\documentstyle [aps,epsf]{revtex}

\def \kaione {{\tilde \chi}_1^0}
\def \kaioneplus {{\tilde \chi}_1^+}
\def \kaitwo {{\tilde \chi}_2^0}

\def \kaioneplus {{\tilde \chi}^+_1}

\def \kaiplusi {{\tilde \chi}^+_i}

\def\Gravitino{\tilde{G}}

\def\stop{\tilde{t}}

\def\pb-1{pb^{-1}}
\def\ppbar{{\bar p}p}
\newcommand{\met}{\mbox{${\not\! E}_{t}$}}
\def\eeggmet{ee\gamma\gamma\met}
\newcommand{\mett}{\mbox{${\not\! E}_{\rm t}$}}
\def\goes{\rightarrow}
\def\pbarp{{\bar p}p}
\def\roots{{\sqrt s}}
\newcommand{\Zee}{\mbox{$Z \rightarrow e^+e^-$}}
\def\ccbar{c{\bar c}}

\begin{document}
\draft

\title{ 
Searches for New Physics in Events with a Photon and $b$--quark Jet
at CDF}

\date{\today}

\maketitle

\font\eightit=cmti8
\def\r#1{\ignorespaces $^{#1}$}
\hfilneg
\begin{sloppypar}
\noindent
T.~Affolder,\r {23} H.~Akimoto,\r {45}
A.~Akopian,\r {37} M.~G.~Albrow,\r {11} P.~Amaral,\r 8  
D.~Amidei,\r {25} K.~Anikeev,\r {24} J.~Antos,\r 1 
G.~Apollinari,\r {11} T.~Arisawa,\r {45}
% ...
\end{sloppypar}
\vskip .026in
\begin{center}
(CDF Collaboration)
\end{center}

\vskip .026in
\begin{center}
\r 1  {\eightit Institute of Physics, Academia Sinica, Taipei, Taiwan 11529, 
Republic of China} \\
\r 2  {\eightit Argonne National Laboratory, Argonne, Illinois 60439} \\
\r 3  {\eightit Istituto Nazionale di Fisica Nucleare, University of Bologna,
I-40127 Bologna, Italy} \\
\r 4  {\eightit Brandeis University, Waltham, Massachusetts 02254} \\
\r 5  {\eightit University of California at Davis, Davis, California  95616} \\
\r 6  {\eightit University of California at Los Angeles, Los 
Angeles, California  90024} \\  
\r 7  {\eightit Instituto de Fisica de Cantabria, CSIC-University of Cantabria, 
39005 Santander, Spain} \\
\r 8  {\eightit Enrico Fermi Institute, University of Chicago, Chicago, 
Illinois 60637} \\
\r 9  {\eightit Joint Institute for Nuclear Research, RU-141980 Dubna, Russia}
\\
\r {10} {\eightit Duke University, Durham, North Carolina  27708} \\
\r {11} {\eightit Fermi National Accelerator Laboratory, Batavia, Illinois 
60510} \\
\r {12} {\eightit University of Florida, Gainesville, Florida  32611} \\
\r {13} {\eightit Laboratori Nazionali di Frascati, Istituto Nazionale di Fisica
               Nucleare, I-00044 Frascati, Italy} \\
\r {14} {\eightit University of Geneva, CH-1211 Geneva 4, Switzerland} \\
\r {15} {\eightit Glasgow University, Glasgow G12 8QQ, United Kingdom}\\
\r {16} {\eightit Harvard University, Cambridge, Massachusetts 02138} \\
\r {17} {\eightit Hiroshima University, Higashi-Hiroshima 724, Japan} \\
\r {18} {\eightit University of Illinois, Urbana, Illinois 61801} \\
\r {19} {\eightit The Johns Hopkins University, Baltimore, Maryland 21218} \\
\r {20} {\eightit Institut f\"{u}r Experimentelle Kernphysik, 
Universit\"{a}t Karlsruhe, 76128 Karlsruhe, Germany} \\
\r {21} {\eightit Center for High Energy Physics: Kyungpook National
University, Taegu 702-701; Seoul National University, Seoul 151-742; and
SungKyunKwan University, Suwon 440-746; Korea} \\
\r {22} {\eightit High Energy Accelerator Research Organization (KEK), Tsukuba, 
Ibaraki 305, Japan} \\
\r {23} {\eightit Ernest Orlando Lawrence Berkeley National Laboratory, 
Berkeley, California 94720} \\
\r {24} {\eightit Massachusetts Institute of Technology, Cambridge,
Massachusetts  02139} \\   
\r {25} {\eightit University of Michigan, Ann Arbor, Michigan 48109} \\
\r {26} {\eightit Michigan State University, East Lansing, Michigan  48824} \\
\r {27} {\eightit University of New Mexico, Albuquerque, New Mexico 87131} \\
\r {28} {\eightit The Ohio State University, Columbus, Ohio  43210} \\
\r {29} {\eightit Osaka City University, Osaka 588, Japan} \\
\r {30} {\eightit University of Oxford, Oxford OX1 3RH, United Kingdom} \\
\r {31} {\eightit Universita di Padova, Istituto Nazionale di Fisica 
          Nucleare, Sezione di Padova, I-35131 Padova, Italy} \\
\r {32} {\eightit University of Pennsylvania, Philadelphia, 
        Pennsylvania 19104} \\   
\r {33} {\eightit Istituto Nazionale di Fisica Nucleare, University and Scuola
               Normale Superiore of Pisa, I-56100 Pisa, Italy} \\
\r {34} {\eightit University of Pittsburgh, Pittsburgh, Pennsylvania 15260} \\
\r {35} {\eightit Purdue University, West Lafayette, Indiana 47907} \\
\r {36} {\eightit University of Rochester, Rochester, New York 14627} \\
\r {37} {\eightit Rockefeller University, New York, New York 10021} \\
\r {38} {\eightit Rutgers University, Piscataway, New Jersey 08855} \\
\r {39} {\eightit Texas A\&M University, College Station, Texas 77843} \\
\r {40} {\eightit Texas Tech University, Lubbock, Texas 79409} \\
\r {41} {\eightit Institute of Particle Physics, University of Toronto, Toronto
M5S 1A7, Canada} \\
\r {42} {\eightit Istituto Nazionale di Fisica Nucleare, University of Trieste/
Udine, Italy} \\
\r {43} {\eightit University of Tsukuba, Tsukuba, Ibaraki 305, Japan} \\
\r {44} {\eightit Tufts University, Medford, Massachusetts 02155} \\
\r {45} {\eightit Waseda University, Tokyo 169, Japan} \\
\r {46} {\eightit University of Wisconsin, Madison, Wisconsin 53706} \\
\r {47} {\eightit Yale University, New Haven, Connecticut 06520} \\
\r {(\ast)} {\eightit Now at Carnegie Mellon University, Pittsburgh,
Pennsylvania  15213} \\
\r {(\ast\ast)} {\eightit Now at Northwestern University, Evanston, Illinois 
60208}
\end{center}

\newpage

\begin{abstract}
We have searched for evidence of physics beyond the standard model in
events that include an energetic photon and an energetic $b$-quark jet,
produced in 85~pb$^{-1}$ of $\ppbar$ collisions at 1.8 TeV at the
Tevatron Collider at Fermilab.  This signature, containing at least
one gauge boson and a third-generation quark, could arise in the
production and decay of a pair of new particles, such as those
predicted by Supersymmetry, leading to a production rate exceeding
standard model predictions. We also search these events for anomalous
production of missing transverse energy, additional jets and leptons
(e, $\mu$ and $\tau$), and additional $b$-quarks.  We find no evidence
for any anomalous production of $\gamma b$ or $\gamma b+X$ events.  We
present limits on two supersymmetric models: a model where the
photon is produced in the decay ${{\tilde\chi}^0_2}\rightarrow\gamma
{{\tilde\chi}^0_1}$, and a model where the photon is
produced in the neutralino decay into the Gravitino LSP,
${{\tilde\chi}^0_1}\rightarrow \gamma \tilde G$.  
We also present our
limits in a model--independent form and test methods of applying
model--independent limits.
\end{abstract}


\section{Introduction} 

As the world's highest--energy accelerator, the Tevatron Collider
provides a unique opportunity to search for evidence of physics beyond
the standard model. There are many possible additions to the standard
model, such as extra spatial dimensions, additional quark generations,
additional gauge bosons, quark and lepton substructure, weak--scale
gravitational effects, new strong forces, and/or supersymmetry, which
may be accessible at the TeV mass scale. In addition, the source of
electro-weak symmetry breaking, also below this mass scale, could well
be more complicated than the standard model Higgs mechanism.

New physics processes are expected to involve the production of heavy
particles,which can decay into standard model constituents (quarks,
gluons, and electroweak bosons) which in turn decay to hadrons and
leptons. Due to the large mass of the new parent particles, the decay
products will be observed with large momentum transverse to the beam
($p_t$), where the rate for standard model particle production is
suppressed. In addition, in many models these hypothetical particles
have large branching ratios into photons, leptons, heavy quarks or
neutral non-interacting particles, which are relatively rare at large
values of $p_t$ in ordinary proton-antiproton collisions.

In this paper we present a broad search for phenomena beyond those
expected in the standard model by measuring the production rate of
events containing at least one gauge boson, in this case the photon,
and a third-generation quark, the b-quark, both with and without
additional characteristics such as missing transverse energy
($\met$).  Accompanying searches are made within these samples for
anomalous production of jets, leptons, and additional b-quarks, which
are predicted in models of new physics.  In addition,
the signature of one gauge-boson plus a third-generation
quark is rare in the standard model, and thus provides an excellent
channel in which to search for new production mechanisms.

The initial motivation of this analysis was a search for the stop squark 
($\stop$) stemming from the unusual $\eeggmet$ 
event observed at the Collider Detector at Fermilab (CDF)\cite{diphoton}.  
A model was proposed\cite{gktop} that produces the photon from the
radiative decay of the $\kaitwo$ neutralino, selected to be the photino, into
the $\kaione$, selected to be the orthogonal state of purely higgsino, and a
photon.  
The production of a chargino-neutralino pair, $\kaiplusi\kaitwo$,
could produce the $\gamma b \mett$ final state via the decay chain

\begin{equation}
\kaiplusi\kaitwo \goes(\stop b)(\gamma\kaione) \goes
(bc\kaione)(\gamma\kaione).
\end{equation}

This model, however, represents only a small part of the available
parameter space for models of new physics. Technicolor models,
supersymmetric models in which supersymmetry is broken by gauge
interactions, models of new heavy quarks, and models of compositness
predicting an excited $b$ quark which decays to $\gamma b$, 
for example, would also create this signature.  
We have consequently generalized the search,
emphasizing the signature ($\gamma b $ or $\gamma b \mett$) rather
than this specific model.  We present generalized, model--independent
limits. Ideally, these generic limits could be applied to actual
models of new physics to provide the information on whether models are
excluded or allowed by the data.  Other procedures for
signature--based limits have been presented
recently\cite{diphoton,cargese,sleuth}.

In the next section we begin with a description of the data selection  followed
by a description of the calculation of backgrounds and  observations of the
data.  Next we present rigorously--derived limits on both Minimal 
Supersymmetric (MSSM) and 
Gauge-mediated Supersymmetry Breaking (GMSB) Models.
The next sections present  the model--independent limits.
Finally, in the Appendix we present tests of the
application of model--independent limits to a variety of models that
generate this signature.

A search for the heavy Techniomega, $\omega_T$, in the final state
$\gamma+b+jet$, derived from the same data sample, has already been 
published\cite{technipaper}. 

\section{Data Selection} 
\label{section:dataselection}

The data used here correspond to  85 pb$^{-1}$ of $\pbarp$
collisions at $\roots=1.8$ TeV. The data sample was collected by triggering on
the electromagnetic  cluster caused by the photon in the central calorimeter. 
We use `standard' photon identification cuts developed for previous photon
analyses~\cite{diphoton},  which are similar to standard electron  requirements
except that there is a restriction on any tracks near the cluster.  The events
are required to have at least one jet with a secondary vertex found by the 
standard silicon detector $b$--quark identification algorithm. 
Finally, we apply missing transverse energy requirements and other selections 
to examine subsamples.
We discuss the selection in detail below.

\subsection{The CDF Detector} 

We briefly describe the relevant aspects of the CDF
detector~\cite{detector}.  A superconducting solenoidal magnet
provides a 1.4~T magnetic field in a volume 3~m in diameter and 5~m
long, containing three tracking devices.  Closest to the beamline is a
4-layer silicon microstrip detector (SVX)~\cite{svxnim} used to
identify the secondary vertices from $b$--hadron decays.  A track
reconstructed in the SVX has an impact parameter resolution of
19$\mu$m at high momentum to approximately 25$\mu$m at 2~GeV/c of
track momentum.  Outside the SVX, a time projection chamber (VTX)
locates the $z$ position of the interaction.  In the region with
radius from 30~cm to 132~cm, the central tracking chamber (CTC)
measures charged--particle momenta.  Surrounding the magnet coil is
the electromagnetic calorimeter, which is in turn surrounded by the
hadronic calorimeter.  These calorimeters are constructed of towers,
subtending $15^\circ$ in $\phi$ and 0.1 in $\eta$\cite{coo}, pointing
to the interaction region.  The central preradiator wire chamber (CPR)
is located on the inner face of the calorimeter in the central region
(\mbox{$|\eta|<1.1$}).  This device is used to determine if the origin
of an electromagnetic shower from a photon was in the magnet coil.  At
a depth of six radiation lengths into the electromagnetic calorimeter
(and 184 cm from the beamline), wire chambers with additional cathode
strip readout (central electromagnetic strip chambers, CES) measure two
orthogonal profiles of showers. 

For convenience we report all energies in GeV, all momenta as
momentum times $c$ in GeV, and all masses as mass times $c^2$ in GeV.
Transverse energy ($E_t$) is the energy deposited in the calorimeter 
multiplied by $\sin\theta$.

\subsection{Event Selection} 

Collisions that produce a photon candidate are selected by at least 
one of a pair of three-level triggers, each of which requires a central
electromagnetic cluster.  The dominant high--$E_t$ photon
trigger requires a 23~GeV cluster with less than approximately 5~GeV
additional energy in the region of the calorimeter surrounding the
cluster~\cite{trigger}. A second trigger, designed to have high
efficiency at large values of $E_t$, requires a 50~GeV cluster, but
has no requirement on the isolation energy.

These events are required to have no energy deposited in the hadronic
calorimeter outside of the time window that corresponds to the beam
crossing.  This rejects events where the electromagnetic cluster was
caused by a cosmic ray muon which scatters and emits bremsstrahlung in
the calorimeter.  

Primary vertices for the $\ppbar$ collisions are 
reconstructed in the VTX system.  A primary vertex is selected as the 
one with the largest total $|p_t|$ attached to it, followed by adding 
silicon tracks for greater precision.
This vertex is required to be less
than 60~cm from the center of the detector along the beamline, so that
the jet is well-contained and the projective nature of the
calorimeters is preserved.

\subsection{Photon} 

To purify the photon sample in the offline analysis, we select events
with an electromagnetic cluster with $E_t>$25~GeV and $|\eta|<1.0$.
To provide for a reliable energy measurement we require the cluster to
be away from cracks in the calorimeter.  To remove backgrounds from
jets and electrons, we require the electromagnetic cluster to be
isolated.  Specifically, we require that the shower shape in the CES
chambers at shower maximum be consistent with that of a single
photon, that there are no other clusters nearby in the CES, and that
there is little energy in the hadronic calorimeter towers associated
with ({\it i.e.} directly behind) the electromagnetic towers of the cluster.

We allow  no tracks with  $p_t>1$~GeV to point at the cluster, and  at most one
track with $p_t<1$~GeV. We require that the sum of the $p_t$ of all tracks
within a cone of $\Delta R =\sqrt{\Delta\eta^2+\Delta\phi^2}=0.4$  around the
cluster be less than 5 GeV.

If the photon cluster has $E_t<$~50~GeV, we require the energy
in a $3\times 3$ array of trigger towers (trigger towers are made of 
two consecutive physical towers in $\eta$) to be less than 4~GeV.
This isolation energy sum excludes the energy in the  
electromagnetic calorimeter trigger tower with the largest energy.
This requirement is more restrictive than the hardware 
trigger isolation requirement,
which is approximately 5~GeV on the same quantity.
In some cases the photon shower leaks into adjacent towers 
and the leaked photon shower energy is included in the isolation energy sum.
This effect leads to an approximately 20\% inefficiency for this trigger. 
When the cluster $E_t$ is above 50~GeV, a second trigger 
with no isolation requirement accepts the event.
For these events, we require the 
transverse energy found in the calorimeter in a cone of $R=0.4$ around 
the cluster to be less than 10\% of the cluster's energy.

These requirements yield a data sample of  511,335 events 
in an exposure of 85~${\rm pb}^{-1}$ of integrated luminosity.

\subsection{B-quark Identification} 

Jets in the events are clustered with a cone of 0.4 in $\eta-\phi$
space using the standard CDF algorithm\cite{jets}.  One of the jets
with $|\eta|<2$ is required to be identified as a $b$--quark jet by
the displaced-vertex algorithm used in the top--quark
analysis\cite{top}.  This algorithm searches for tracks in the SVX
that are associated with the jet but not associated with the primary
vertex, indicating they come from the decay of a long--lived particle.
We require that the track, extrapolated to the interaction vertex, has
a distance of closest approach greater than 2.5 times its uncertainty
and pass loose requirements on $p_t$ and hit quality.  The tracks
passing these cuts are used to search for a vertex with three or more
tracks.  If no vertex is found, additional requirements are placed on
the tracks, and this new list is used to search for a two--track
vertex.  The transverse decay length, $L_{xy}$, is defined in the
transverse plane as the projection of the vector pointing from the
primary vertex to the secondary vertex on a unit vector along the jet
axis.  We require $|L_{xy}|/\sigma >3$, where $\sigma$ is the
uncertainty on $L_{xy}$.  These requirements constitute a ``tag''.  In
the data sample the tag is required to be positive, with $L_{xy}>0$.
The photon cluster can have tracks accidentally associated with it and
could possibly be tagged; we remove these events.  This selection
reduces the dataset to 1487 events.

The jet energies are corrected for calorimeter gaps and non-linear
response, energy not contained in the jet cone, and underlying event
energy\cite{jets}.  For each jet the resulting corrected $E_t$ is the
best estimate of the underlying true quark or gluon transverse energy,
and is used for all jet requirements in this analysis.  We require the
$E_t$ of the tagged jet in the initial $b\gamma$ event selection to be greater
than 30~GeV; this reduces the data set to 1175 events.

\subsection{Other Event Selection} 
\label{section:otherselection}

While the photon and $b$--tagged jet constitute the core of the
signature we investigate, supersymmetry and other new physics could be
manifested in any number of different signatures.  Because of the
strong dependence of signature on the many parameters in
supersymmetry, one signature is (arguably) not obviously more likely
than any other.  For these reasons we search for events with unusual
properties such as very large missing $E_t$ or additional
reconstructed objects.  These objects may be jets, leptons, additional
photons or $b$-tags.  This method of sifting events was employed in a
previous analysis\cite{diphoton}.  We restrict ourselves to objects
with large $E_t$ since this process is serving as a sieve of the
events for obvious anomalies.  In addition, in the lower $E_t$ regime
the backgrounds are larger and more difficult to calculate.  In this
section we summarize the requirements that define these objects.

Missing $E_t$ ($\met$) is the magnitude 
of negative of the two-dimensional vector
sum of the measured $E_t$ in each calorimeter tower with energy above
a low threshold in the region $|\eta| < 3.6$.  All jets in the event
with uncorrected $E_t$ greater than 5~GeV and $|\eta|<$2 are corrected
appropriately for known systematic detector mismeasurements; these
corrections are propagated into the missing $E_t$.  Missing $E_t$ is
also corrected using the measured momentum of muons, which do not
deposit much of their energy in the calorimeter.

We apply a requirement of 20~GeV on missing $E_t$, and observe that a
common topology of the events is a photon opposite in azimuth from the
missing $E_t$ (see Figure \ref{figoptnb}).  We conclude that a common
source of missing $E_t$ occurs when the basic event topology is a
photon recoiling against a jet.  This topology is likely to be
selected by the $\met$ cut because the fluctuations in the measurement
of jet energy favor small jet energy over large.  To remove this
background, we remove events in the angular bin  
$\Delta\phi(\gamma-\met)>168^\circ$ for the
sample, where we have raised the missing $E_t$ requirement to 40~GeV.

We define $H_t$ as the scalar sum of the $E_t$ in the calorimeter 
added to the missing $E_t$ and the $p_t$ of any muons in the event.
This would serve as a measure of the mass scale of new objects
that might be produced.

To be recognized as an additional jet in the event, a calorimeter
cluster must have corrected $E_t>$15~GeV and $|\eta|<2$.  To count as
an additional $b$ tag, a jet must be identified as a $b$ candidate by
the same algorithm as the primary $b$ jet, and have $E_t>30$~GeV and
$|\eta|<$2.
To be counted as an additional photon, an electromagnetic 
cluster is required to have $E_t>25$~GeV, $|\eta|<1.0$, and to pass all the 
same identification requirements as the primary photon.

\begin{table}
\begin{center}
\begin{tabular}{|l|l|} 
Object & Selection \\ \hline
\multicolumn{2}{|c|}{Basic Sample Requirements} \\ \hline
Isolated Photon & $E_t$ $>$ 25 GeV, $|\eta|<1.0$ \\
$b$-quark jet (SVX $b$-tag)    & $E_t$ $>$ 30 GeV, $|\eta|<2.0$  \\
\hline
\multicolumn{2}{|c|}{Optional Missing $E_t$ Requirements} \\
\hline
$\met$                     & $>$ 40 GeV \\
$|\Delta \phi(\gamma -\met)|$ & $< 168^{\circ}$ \\
\hline
\multicolumn{2}{|c|}{Optional Other Objects} \\
\hline
Jets                    & $E_t >$ 15 GeV, $|\eta|<2.0$ \\
Additional Photons      & $E_t >$ 25 GeV, $|\eta|<1.0$ \\
Additional $b$-quark jets & $E_t >$ 30 GeV, SVX $b$-tag \\
Electrons               & $E_t >$ 25 GeV, $|\eta|<1.0$ \\
Muons                   & $p_t >$ 25 GeV, $|\eta|<1.0$ \\
Tau Leptons             & $E_t >$ 25 GeV, $|\eta|<1.2$\\
\end{tabular}
\end{center}
\caption{Summary of the kinematic selection criteria for the 
$b\gamma+X$ sample that contains 1175 events.  Also shown are the
kinematic criteria for the identification of other objects, such as
missing $E_t$, jets, additional b-jets and leptons. The lepton
identification criteria are the same as used in the top discovery.}
\label{tab:selection}
\end{table}

For lepton identification, we use the cuts defined for the primary
leptons in the top quark searches\cite{top,tau}.  We search for
electrons in the central calorimeter and for muons in the central muon
detectors.  Candidates for $\tau$ leptons are identified only by their
hadronic decays -- as a jet with one or three high--$p_t$ charged
tracks, isolated from other tracks and with calorimeter energy cluster
shapes consistent with the $\tau$ hypothesis\cite{tau}.  
Electrons and
$\tau$'s must have $E_t>25$~GeV as measured in the calorimeter; muons
must have $p_t>25$~GeV.  Electrons and muons must have $|\eta|<1.0$
while $\tau$'s must have $|\eta|<1.2$.
We summarize the kinematic selections in Table~\ref{tab:selection}.

\section{Background Estimates} 
\label{section:primary}

The backgrounds to the $b\gamma$ sample are combinations of standard
model production of photons and $b$ quarks and also jets misidentified
as a photon (``fake'' photons) or as a $b$--quark jet (``fake'' tags
or mistags).  A jet may be misidentified as a photon by fragmenting to
a hard leading $\pi^0$.  Other jets may fake a $b$--quark jet through
simple mismeasurement of the tracks leading to a false secondary
vertex.

We list these backgrounds and a few other smaller backgrounds in Table
\ref{tab:bgmethods}.  The methods referred to in this table are
explained in the following sections.

\begin{table}[!ht]
\centering
\begin{tabular}{|l|l|}
Source & Method of Calculation \\ \hline
$\gamma b\bar{b}$ and $\gamma c\bar{c}$  & Monte Carlo \\ \hline 
$\gamma+$ mistag  & CES--CPR and tagging prediction\\ \hline 
fake $\gamma$ and $b\bar{b}$ or $c\bar{c}$  & CES--CPR \\ \hline 
fake $\gamma$ and a mistag  & CES--CPR \\ \hline 
$W\gamma$, $Z\gamma$  & Monte Carlo, normalized to data\\ \hline 
electrons faking $\gamma$'s  & measured fake rate \\ \hline 
cosmic rays  & cosmic characteristics \\ 
\end{tabular}
\caption{The summary of the backgrounds to the photon and tag sample
and the methods used to calculate them.}
\label{tab:bgmethods}
\end{table}

The following sections begin with a discussion of the tools used to
calculate backgrounds.  Section \ref{section:bgmethod} explains why
the method presented is necessary.  The subsequent sections provide
details of the calculation of each background in turn.

\subsection{Photon Background Tools} 
\label{section:phoback}

There are two methods we use to calculate photon backgrounds, each
used in a different energy region.  The first employs the CES detector
embedded at shower maximum in the central electromagnetic
calorimeter\cite{photonbg}.  This method is based on the fact that the
two adjacent photons from a high--$p_t$ $\pi^0$ will tend to
create a wide CES cluster, with a larger CES $\chi^2$, when compared
to the single photon expectation.  The method produces an
event-by-event weight based on the $\chi^2$ of the cluster and the
respective probabilities to find this $\chi^2$ for a $\pi^0$
versus for a photon.  In the decay of very high--energy $\pi^0$'s
the two photons will overlap, and the $\pi^0$ will become
indistinguishable from a single photon in the CES by the shape of the
cluster. From studies of $\pi^0$'s from $\rho$ decay we have found
that for $E_t>35$~GeV the two photons coalesce and we must use a
second method of discrimination that relies on the central preradiator
system (CPR)\cite{photonbg}.  This background estimator is based on
the fact that the two photons from a $\pi^0$ have two chances to
convert to an electron-positron pair at a radius before the CPR
system, versus only one chance for a single photon. The charged
particles from the conversion leave energy in the CPR, while an
unconverted photon does not.  The implementation of the CPR method of
discriminating photons from $\pi^0$'s on a statistical basis is
similar to the CES method, an event-by-event weight.  When the two
methods are used together to cover the entire photon $E_t$ range for a
sample, we refer to it as the CES--CPR method.

Both these photon background methods have low discrimination power at
high photon $E_t$.  This occurs because the weights for a single
photon and a (background) $\pi^0$ are not very different.  For
example, in the CES method, at an $E_t$ of 25~GeV, the probability for
a photon to have a large $\chi^2$ is on the order of 20\% while the
background has a probability of approximately 45\%.
For the CPR method, typical values for a 25~GeV photon 
are 83\% conversion probability for background and 60\% for a single photon.

\subsection{$b$-quark Tagging Background Tools} 

A control sample of QCD multi-jet events is used to study the
backgrounds to the identification of $b$--quark jets\cite{topback}.
For each jet in this sample, the $E_t$ of the jet, the number of SVX
tracks associated with the jet, and the scalar sum of the $E_t$ in the
event are recorded.  The probability of tagging the jet is determined
as a function of these variables for both positive ($L_{xy}>0$) and
negative tags ($L_{xy}<0$).

Negative tags occur due to measurement resolution and errors in
reconstruction.  Since these effects produce negative and positive
tags with equal probability, the negative tagging probability can be
used as the probability of finding a positive tag due to
mismeasurement (mistags).

\subsection{Background Method} 
\label{section:bgmethod}

We construct a total background estimate from summing the individual
sources of backgrounds, each found by different methods.  
In the CDF top analysis \cite{top} one of the tagging background 
procedures was to apply the positive tagging probability to the 
jets in the untagged sample to arrive at a total 
tagging background estimate.
A similar procedure could be considered for our sample.

However, in this analysis, 
a more complex background calculation is necessary for two reasons.
First, the parameterized tagging background described above is derived from 
a sample of jets from QCD events\cite{top} which
have a different fraction of $b$--quark jets than do jets in 
a photon-plus-jets sample.  This effect is caused by the 
coupling of the photon to the quark charge.
Secondly, $b$ quarks produce $B$ mesons which have a large
branching ratio to semileptonic states that include neutrinos, 
producing real missing $E_t$ more often than generic jets.
When a $\met$ cut is applied,
the $b$ fraction tends to increase.  This effect is averaged over in the
positive background parameterization so the background prediction 
will tend to be high at small $\met$ and low at large $\met$.

For these reasons, the positive tagging rate is correlated to the 
existence of a photon and also the missing $E_t$, when that is required.
In contrast, the negative tagging rate is found not to be 
significantly correlated with the presence of real $b$ quarks.
This is because the negative tagging rate is due only to mismeasurement
of charged tracks which should not be sensitive to the 
flavor of the quarks.

The next sections list the details of the calculations of the 
individual sources of the backgrounds.  
Both photons and $b$--tagged jets have significant backgrounds so 
we consider sources with real photons and $b$--tags 
or jets misidentified as photons or $b$--jets (``fakes'').

\subsection{Heavy Flavor Monte Carlo} 

The background consisting of correctly--identified photons and
$b$--quark jets is computed with an absolutely normalized Monte Carlo
\cite{MLM}. The calculation is leading order, based on $q\bar q$ and $gg$
initial states and a finite $b$--quark mass.  The $Q^2$ scale is taken
to be the
square of the photon $E_t$ plus the square of the $b\bar{b}$ or $c\bar{c}$ 
pair mass, $Q^2=E_t^2+M^2$.
A systematic uncertainty of 30\% is found by scaling $Q$ by a factor of two
and the quark masses by 10\%.  
An additional 20\% uncertainty allows for additional effects 
which cannot be determined 
by simply changing the scale dependence\cite{MLM}.

In addition, we rely on the detector simulation of the Monte Carlo to
predict the tail of the rapidly falling $\met$ spectrum.  The Monte
Carlo does not always predict this tail well.  For example, a Monte
Carlo of $\Zee$ production predicts only half the observed rate for events
passing the missing $E_t$ cut used in this analysis. We thus include an
uncertainty of 100\% on the rate that events in the $b\gamma$ sample
pass the $\met$ cut.  We combine the Monte Carlo production and $\met$
sources of uncertainty in quadrature.  However when the $\gamma
b\bar{b}$ and $\gamma c\bar{c}$ backgrounds are totaled, these common
uncertainties are treated as completely correlated.

\subsection{Fake photons} 

The total of all backgrounds with fake photons can be measured using
the CES and CPR detectors.  These backgrounds, dominated by jets that
fragment to an energetic $\pi^0\rightarrow \gamma\gamma$ and 
consequently are misidentified as a single photon, are measured using the
shower shape in the CES system for photon $E_t<35$~GeV and the
probability of a conversion before the CPR for
$E_t>35$~GeV~\cite{photons}.  We find $55\pm 1\pm 15\%$~\cite{uncert}
of these photon candidates are actually jets misidentified as photons.

For many of our subsamples we find this method is not useful
due to the large statistical dilution as explained in Section 
\ref{section:primary}.  This occurs because,
for example, the probabilities for background ($\pi^0$'s) and for
signal ($\gamma$'s) to convert before the CPR are not too different.
This results in a weak separation and a poor statistical uncertainty.
We find the method returns 100\% 
statistical uncertainties for samples of less
than approximately 25 photon candidates.

\subsection{Real photon, Fake tags} 

To estimate this background we start with the untagged sample, and weight it
with both the CES-CPR real photon weight and the negative 
tagging (background) weight.
This results in the number of true photons 
with mistags in the final sample.  
As discussed above, the negative tagging prediction does not have 
the correlation to quark flavor and missing $E_t$ as does 
the positive tagging prediction. 

As a check, we can look at the sample before the tagging and $\met$
requirements.  In this sample we find 197 negative tags while the
estimate from the negative tagging prediction is 312.  This
discrepancy could be due to the topology of the events -- unlike
generic jets, the photon provides no tracks to help define the primary
vertex.  The primary vertex could be systematically mismeasured
leading to mismeasurement of the transverse decay length $L_{xy}$ for
some events.  We include a 50\% uncertainty on this background due to
this effect.

\subsection{Estimate of Remaining Backgrounds} 

There are several additional backgrounds which we have calculated and
found to very small.  The production of $W\gamma$ and $Z\gamma$ events
may provide background events since they produce real photons and $b$
or $c$ quarks from the boson decay ($W^\pm\rightarrow c\bar{s}$,
$Z\rightarrow b\bar{b}$).  The $\met$ would have to be fake, due to
mismeasurement in the calorimeter.  We find $W/Z\gamma$ events in the
CDF data using the same photon requirements as the search.  The $W/Z$
is required to decay leptonically for good identification.  We then
use a Monte Carlo to measure the ratio of the number of these events
to the number of events passing the full $\gamma b \met$ search cuts.
The product of these two numbers predicts this background to be less
than 0.1 events.

The next small background is $W\rightarrow e\nu$ plus jets where the
electron track is not reconstructed, due either to bremsstrahlung or
to pattern-recognition failure.  Using $\Zee$ events, we find
this probability is small, about 0.5\%.  Applying this rate to the
number of observed events with an electron, $b$-tag and missing $E_t$
we find the number of events expected in our sample to be negligible.

The last small background calculation is the rate for cosmic ray events.
In this case there would have to be a QCD $b$--quark event with a cosmic ray 
bremsstrahlung in time with the event.  The missing $E_t$ comes with the 
unbalanced energy deposited by the cosmic ray.
We use the probability that a cosmic ray leaves an unattached stub in the 
muon chambers to estimate that the number of events in this category is also
negligible.

The total of all background sources is summarized in
Table~\ref{tabmtwo}.  The number of observed events is consistent
with the calculation of the background for both the $\gamma b$ sample
and the subsamples with $\met$.

\section{Data Observations} 

In this section we report the results of applying the final event
selection to the data.  First we compare the total background estimate
with the observed number of events in the $b\gamma$ sample, which
requires only a photon with $E_t>25$~GeV and a $b$--tagged jet with
$E_t>30$~GeV.  Since most models of supersymmetry predict missing
$E_t$, we also tabulate the backgrounds for that subsample.

Table \ref{tabmtwo} summarizes the data samples 
and the predicted backgrounds.  We find 98 events have missing 
$E_t>20$~GeV.  Six events have missing $E_t>40$~GeV, and 
only two of those events pass the $\Delta\phi(\gamma-\met)<168^\circ$ cut.

\begin{figure}[ht]
\epsfxsize=8.6cm
\hspace*{3.5cm}
\epsfbox{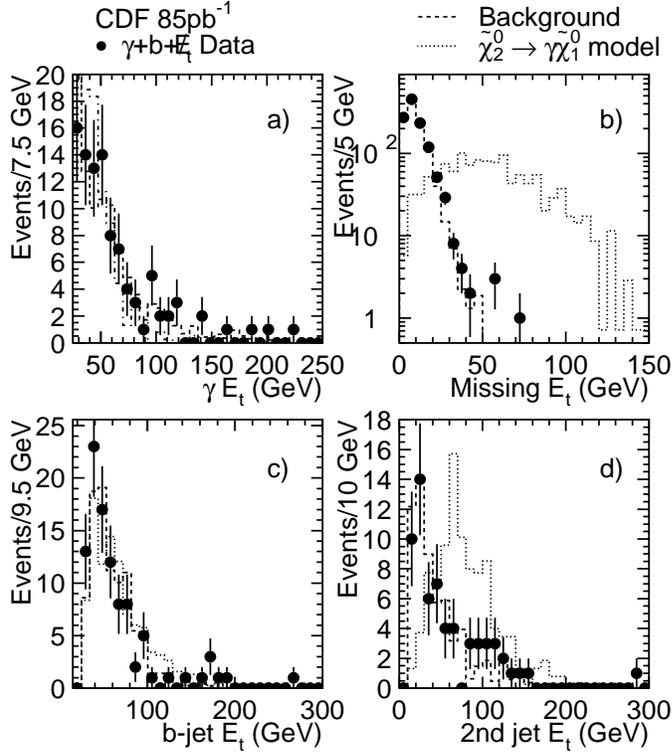}
\caption{Comparison of the data to the background prediction
(dashed line), and the baseline SUSY model of Section
\protect\ref{section:sqandgl} (dotted line).  The data consist of the
98 events of the $\gamma b$ data with $\met>20$~GeV, except in b)
which contains no $\met$ requirement. In each case the predictions
have been normalized to the data.  The distributions are: a) the
photon $E_t$, b) the missing $E_t$, c) the $b$--tagged jet $E_t$ and
d) the $E_t$ of the second jet with $E_t>$15~GeV, if there is one.
For display, 
the SUSY model event yield is scaled up by a factor of 4 for a), c)
and d) and a factor of 40 for b).}
\label{figoptna}
\end{figure}

\begin{figure}[ht]
\epsfxsize=8.6cm
\hspace*{3.5cm}
\epsfbox{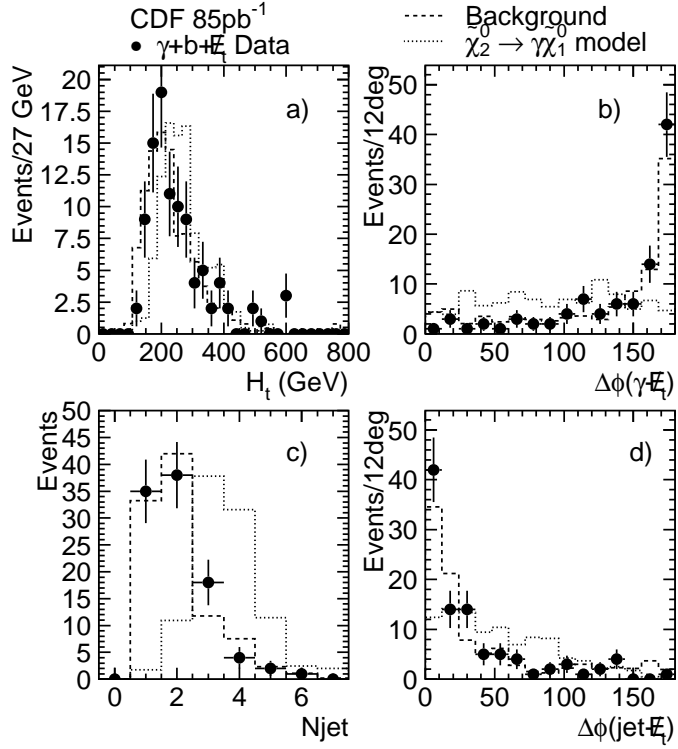}
\caption{Comparison of the data to the background prediction
(dashed line), and the the baseline SUSY model of Section 
\protect\ref{section:sqandgl} (dotted line), 
each normalized to the 98 events of the $\gamma b$ data with
$\met>20$~GeV.  The distributions are: a) $H_t$ (total energy), b)
$\Delta \phi$ between the photon and the $\met$, c) number of jets
with $E_t>15$~GeV, and d) $\Delta\phi$ between the missing $E_t$
and the nearest jet.  
For display, the SUSY model event yield is scaled up by a
factor of 4.}
\label{figoptnb}
\end{figure}

Figures \ref{figoptna} and \ref{figoptnb} display the 
kinematics of the data with a background prediction overlayed.
Because of the large 
statistical uncertainty in the fake photon background, 
the prediction for bins with small statistics have such 
large uncertainties that they are not useful.  
In this case we approximate the fake photon
background by applying the fake photon measurement and the 
positive tagging prediction to the large-statistics 
untagged sample.  This approximation only assumes that
real $b$ quarks do not produce substantial missing $E_t$.
Each component of the background is normalized to the number expected
as shown in Table \ref{tabmtwo};
the total is then normalized to the data in order to compare distributions.
We observe no significant deviations from the expected background.

\begin{table}[ht]
\centering
\begin{tabular}{|c|c|c|c|} 
Source & Events& Events $\met>20$ & Events $\met>40$, $\Delta\phi$\\ \hline
$\gamma b\bar{b}$        &$99\pm 5 \pm 50$     & $9\pm 1 \pm 10$ 
                                   &$0.4\pm 0.3 \pm 0.4$ \\ \hline
$\gamma c\bar{c}$        &$161\pm 9 \pm 81$     & $7\pm 2\pm 8$ 
                                   &$0.0\pm 0.5\pm 0.5$ \\ \hline
$\gamma+$mistag          &$124\pm 1\pm 62$      & $10\pm 0.3\pm 5.2$ 
                                   & $0.7\pm 0.05\pm 0.5$\\ \hline
fake $\gamma$            &$648\pm 69\pm 94$    & $49\pm 22\pm 7$ 
                                   & $1.0\pm 1.0\pm 0.2$\\ \hline
$W\gamma$                & $2\pm 1$            & $0.4\pm 0.2\pm 0.4$ 
                                   & $0.0\pm 0.1\pm 0.1$ \\ \hline
$Z\gamma$                & $6\pm 4$            & $0.8\pm 0.6\pm 0.8$ 
                                   & $0.08\pm 0.06\pm 0.08$ \\ \hline
$e\rightarrow \gamma$    & $0.4\pm0.1$         & $0.4\pm0.1$ 
                                   & $0.1\pm.03$ \\ \hline
cosmics                  & $0\pm 16$           & $0\pm 5$ 
                                   & 0 \\ \hline \hline
total background         & $1040\pm 72\pm 172$   & $77\pm 23\pm 20$ 
                                   & $2.3\pm 1.2\pm 1.1$\\ \hline \hline
data &1175& 98 & 2\\ 
\end{tabular}
\caption{Summary of the primary background calculation.
The $\gamma b\bar{b}$ and $\gamma c\bar{c}$ systematic uncertainties
are considered 100\% correlated.  The column labeled $\met>40$~GeV
also includes the requirement that $\Delta
\phi(\gamma-\met)<168^\circ$.  The entry for fake photons in the
column labeled $\met>40$~GeV is not measured but is estimated using the
assumption that 50\% of photons are fakes.  This number is assigned a
100\% uncertainty.}
\label{tabmtwo}
\end{table}

Several events appear on the tails of some of the distributions.
Since new physics, when it first appears, will likely be at the limit
of our kinematic sensitivity, the tail of any kinematic distribution
is a reasonable place to look for anomalous events.  However, a few
events at the kinematic limit do not warrant much interest unless they
have many characteristics in common or they have additional unusual
properties.  We find two events pass the largest missing $E_t$ cut of
40~GeV; we examine those events more closely below. We also observe
there are five events with large dijet mass combinations and we also
look at those more closely below.  In Section \ref{section:otherobs},
we search for other anomalies in our sample.

\clearpage

\clearpage

\clearpage

\begin{figure}[ht]
\epsfxsize=8.6cm
\hspace*{3.5cm}
\epsfbox{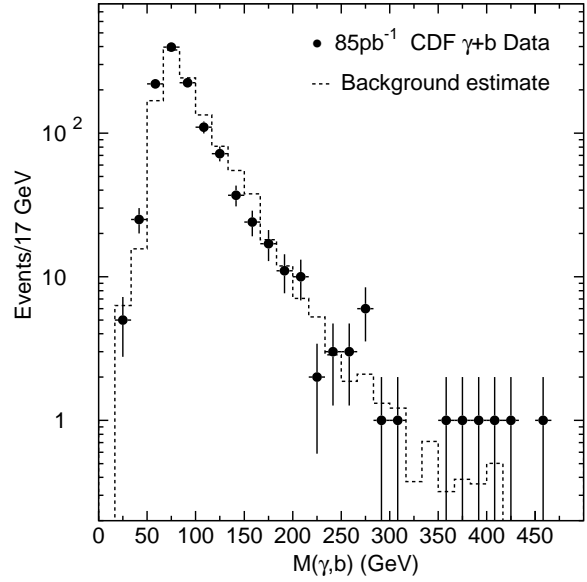}
\caption{Comparison of the $\gamma b$ mass in the data to the 
background prediction
(dashed line), normalized to the 1175 events of the $\gamma b$ data.}
\label{figgammabmass}
\end{figure}

\clearpage
\begin{figure}[ht]
\epsfxsize=8.6cm
\hspace*{3.5cm}
\epsfbox{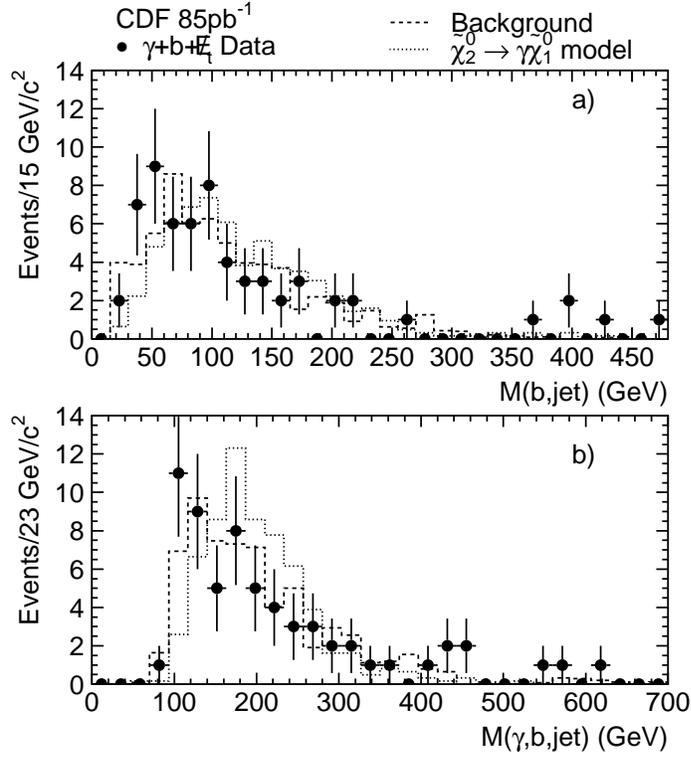}
\caption{The distributions for: a) $M(b,j)$ and b) $M(\gamma,b,j)$ 
for the $\met>20$~GeV events as shown in Figure~\protect\ref{figscatmone}.
Only 63 of the 98 events have a second jet and make it into this plot. 
The data are compared to a background prediction
(dashed line), and the baseline SUSY model of Section 
\protect\ref{section:sqandgl} (dotted line), 
each normalized to the data.  
The Monte Carlo prediction is scaled up by a factor of 3.}
\label{figfiveproj}
\end{figure}

\clearpage

\subsection{Analysis of Events with Large Missing $E_t$} 

Six events pass the {\it a priori} selection criteria requiring a
photon, $b$--tag, and $\met>$40~GeV.  Two of these events also pass
the $\Delta\phi(\gamma-\met)<168^\circ$ requirement.  We have examined
these two events to see if there indications of anything else unusual
about them (for example, a high--$p_t$ lepton, or a second jet 
which forms a large invariant mass with the first $b$-jet, to take
signals of GMSB and Higgsino models respectively).

The first event (67537/59517) does not have the characteristics of a
typical $b$-tag.  It is a two-track tag (which has a worse
signal--to--noise) with the secondary vertex consistent with the beam
pipe radius (typical of an interaction in the beam pipe).  The two
tracks have a $p_t$ of 2 and 60~GeV, respectively; this highly
asymmetric configuration is unlikely if the source is a $b$--jet.
There are several other tracks at the same $\phi$ as the jet that
are inconsistent with either the primary or secondary vertex.  We conclude
the $b$-tag jet in this event is most likely to be a fake, coming
from an interaction in the beam pipe.

The second event has a typical $b$--tag but there are three jets, 
and all three straddle cracks in the calorimeter ($\eta=0.97,-1.19,-0.09$), 
implying the $\met$ is very likely to be mismeasured.

In both events we judge by scanning that the primary vertex is the 
correct choice so that a mismeasurement of the $\met$ due to selecting 
the wrong vertex is unlikely.

While we have scanned these two events and find they are most likely
not true $\gamma b \met$ events, we do not exclude them from the 
event sample as the background calculations include these sources of
mismeasured events.

\begin{table}[ht]
\centering
\begin{tabular}{|c|c|c|c|c|c|c|c|c|} 
Run/Event & $\gamma~E_t$ & \met & $M(b,jet)$ &$b~E_t$ & jets$~E_t$ 
& $\Delta\phi(\gamma-\met)$ & $\Delta\phi(j_{near}-\met)$  & $H_t$ \\ \hline
 60951/189718  & 121 & 42 & 57 & 61 & 67,26,15 & 177 &11& 342 \\ \hline
 64997/119085  & 222 & 44 & 97 & 173 & 47 & 170 &1& 495  \\ \hline
 63684/15166   & 140 & 57 & 63 &  35 & 25,20,15 & 175 &6& 388\\ \hline
 67537/59517$^*$ & 36 & 73 & 399 & 195 & 141,113,46,17 & 124 &20& 595 \\ \hline
 69426/104696  & 33 & 58 & 266 & 143 & 119 & 180 &3& 344 \\ \hline
 68464/291827$^*$ &  93 & 57 & 467 & 128 & 155,69 & 139 &16& 405 \\ 
\end{tabular}
\caption{Characteristics of the six events with $\met>40$~GeV; 
the two marked with an asterisk also pass the 
$\Delta\phi(\gamma-\met)<168^\circ$ requirement.  All units are GeV
except for $\Delta\phi$ which is in degrees.  The columns are 
the $E_t$ of the photon in the event, the missing $E_t$, 
the mass of the $b$-jet and the second highest $E_t$ jet in the event, the
$E_t$ of the $b$ jet, the $E_t$ of the jets other than the $b$ jet, 
the $\Delta\phi$ between the photon and the missing $E_t$, 
the $\Delta\phi$ between the the missing $E_t$ and the nearest jet, 
and the $H_t$
of the event (scalar sum of the $E_t$, the missing $E_t$ and 
the $p_t$ of any muons in the event).}
\label{tabumet}
\end{table}

\subsection{Analysis of Five High--mass Events} 

\begin{figure}[ht]
\epsfxsize=8.6cm
\hspace*{3.5cm}
\epsfbox{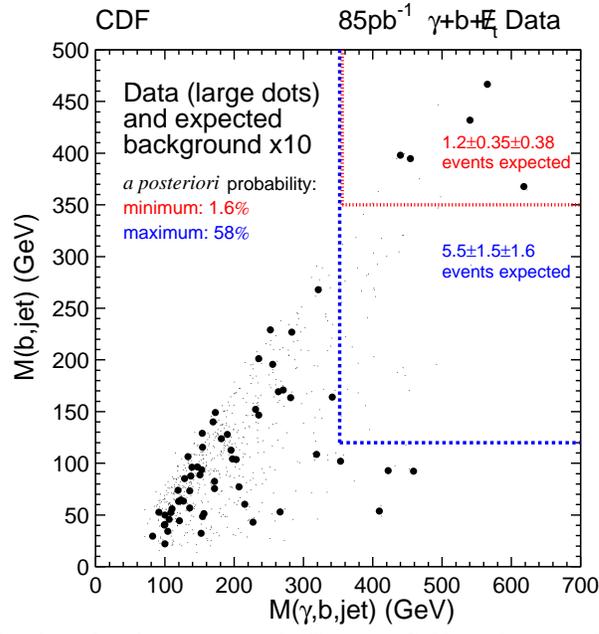}
\caption{$M(b,j)$ versus $M(\gamma,b,j)$ for the events with $\met>20$~GeV 
as shown in Figure \protect\ref{figscatmone}.  Only 63 of the 98 
events have 
a second jet and make it into this plot. The small dots are the
result of making the scatter plot for the untagged data (passing
all other cuts) and weighting it  with the positive tagging prediction.
The estimates of background expected in the boxes are found by 
the method described in the text.}
\label{figscatmone}
\end{figure}

If the events include production of new, heavy particles, 
we might observe peaks, or more likely, 
distortions in the distributions of the 
masses formed from combinations of objects.  To investigate this,
we create a scatter plot of the mass of the $b$-quark jet and the 
second highest-$E_t$ jet versus the mass of the photon, $b$-quark jet
and second highest jet in Figure \ref{figscatmone}
and \ref{figfiveproj}.

As seen in the figures, the five events at highest $M(b,j)$ seem to
form a cluster on the tail of the distribution.  There are 63 events
in the scatter plot which are the subset of the 98 events with
$\met>20$~GeV which contain a second jet.  The five events include the
two (probable background) events with $\met>40$~GeV 
and $\Delta\phi(\gamma-\met)<168^\circ$
and three events
with large $H_t$ ($>400$ GeV).  Since these events were selected for their
high mass, we expect they would appear in the tails of several of the
distributions such as $H_t$.  Table \ref{tabumass} shows the
characteristics of these five events.

\begin{table}[ht]
\centering
\begin{tabular}{|c|c|c|c|c|c|c|c|c|} 
Run/Event & $\gamma~E_t$ & \met & $M(b,jet)$ &b$~E_t$ & jet $E_t$'s 
& $\Delta\phi(\gamma-\met)$ & $\Delta\phi(j_{near}-\met)$  & $H_t$  \\ \hline
  66103/52684  & 106 & 24 & 433 & 170 & 135,57 & 152 &      29& 517\\ \hline
 66347/373704  & 122 & 32 & 369 & 268 & 125,42 & 101 &      14& 605 \\ \hline
  67537/59517  &  36 & 73 & 399 & 195 & 141,113,46,17& 124 &20& 595\\ \hline
 68333/233128  &  38 & 39 & 395 &  99 & 282,212 & 121&       3& 600 \\ \hline
  68464/291827 &  93 & 57 & 467 & 128 & 155,69 & 139 &      16& 405\\ 
\end{tabular}
\caption{Characteristics of events with $M(b,jet)>300$~GeV.
For a complete description of the quantities, see Table \protect\ref{tabumet}.
}
\label{tabumass}
\end{table}

In order to see if these events are significant, we need to make an
estimate of the expected background.  We define the two regions
indicated in Figure \ref{figscatmone}.  The small box is placed so
that it is close to the five events.\footnote{Note that events cannot
be above the diagonal in the $M(\gamma,b,j)-M(b,j)$ plane, so the
true physical area is triangular.}  This is intended to maximize the
significance of the excess.  The large box is placed so that it is as
far from the five events as possible without including any more data
events.  This will minimize the significance.  The two boxes can serve
as informal upper and lower bounds on the significance.  Since these
regions were chosen based on the data, the excess over background
cannot be used to prove the significance of these events.  These
estimates are intended only to give a guideline for the significance.

We cannot estimate the background to these five events using the CES
and CPR methods described in Section~\ref{section:phoback} due to the
large inherent statistical uncertainties in these techniques.  We
instead use the following procedure.  The list of backgrounds in
Section~\ref{section:primary} defines the number of events from each source
with no restriction on $M(b,j)$.  We normalize these numbers to the 63
events in the scatter plot.  We next derive the fraction of each of
these sources we expect at high $M(b,j)$.  We multiply the background
estimates by the fractions.  The result is a background estimate for
the high--mass regions.

To derive the fractions of background sources expected at high
$M(b,j)$ we look at each background in turn.  The fake photons are QCD
events where a jet has fluctuated into mostly electromagnetic energy.
For this source we use the positive $L_{xy}$ background
prediction\cite{top} to provide the fraction.  This prediction is
derived from a QCD jet sample by parameterizing the positive tagging
probability as a function of several jet variables.  The probability
for each jet is summed over all jets for the untagged sample to arrive
at a tagging prediction.  Since the prediction is derived from QCD
jets we expect it to be reliable for these QCD jets also.  Running
this algorithm (called ``Method 1''\cite{top}) on the untagged photon
and $\met$ sample yields the fraction of expected events in each of
the two boxes.  The fractions are summarized in Table
\ref{tab:fivefrac}.

\begin{table}[ht]
\centering
\begin{tabular}{|c|c|c|} 
Source & Big box & Small box  \\ \hline
fake $\gamma$ & $0.080\pm 0.007$ & $0.017\pm 0.003$\\ \hline      
$\gamma$, fake tag & $0.112\pm 0.009$ & $0.032\pm 0.005$\\ \hline
$\gamma b\bar{b}$ & $0.10\pm 0.03$ & $0.022\pm 0.007$\\ \hline
$\gamma c\bar{c}$ & $0.08\pm 0.04$ & $0.018\pm 0.008$\\ 
\end{tabular}
\caption{The fraction of the 63 $\gamma b j\met$ events for
each background expected to fall into the high--$M(b,j)$
boxes defined in Figure \protect\ref{figscatmone}.}
\label{tab:fivefrac}
\end{table}

The second background source considered consists of real photons with
fake tags.  We calculate this contribution using the measured negative
tagging rate applied to all jets (i.e. before $b$-tagging) in the
sample.  Finally, the real photon and heavy flavor backgrounds are
calculated based on the Monte Carlo.  The results from estimating the
fractions are shown in Table
\ref{tab:fivefrac}.

The estimates of the sources of background for the 63 events at all $M(b,j)$
have statistical uncertainties, as do the fractions in 
Table \ref{tab:fivefrac}; 
we include both in the uncertainty in the number of events 
in the high--mass boxes.
We propagate the systematic uncertainties on the backgrounds
to the 63 events at all $M(b,j)$ and include the following systematics 
due to the fractions:
\begin{enumerate}
\item 50\% of the real photon and mistag background calculation 
for the possibility that 
the quark and gluon content, as well as the heavy flavor fraction, 
in photon events may differ from the content in QCD jets.
\item 50\% of the real photon and mistag background calculation
for the possibility that
using the positive tagging prediction to correct
the Monte Carlo for the $\met$ cut may have a bias.
\item 100\% of the real photon and real heavy flavor background 
calculation for the possibility that the tails in the Monte Carlo $M(b,j)$
distribution may not be reliable.
\end{enumerate}
The results of multiplying the backgrounds at all $M(b,j)$
with the fractions expected at high $M(b,j)$ are shown in 
Table \ref{fivetot}.

\begin{table}[ht]
\centering
\begin{tabular}{|c|c|c|} 
Source & Big box & Small box  \\ \hline
fake $\gamma$     & $3.3\pm  1.5 \pm 0.5$  &$0.70\pm 0.33\pm 0.10$\\ \hline
$\gamma$, fake tag& $0.97\pm 0.09\pm 0.69$ &$0.28\pm 0.05\pm 0.20$\\ \hline
$\gamma b\bar{b}$ & $0.75\pm 0.26\pm 1.18$ &$0.16\pm 0.06\pm 0.26$\\ \hline
$\gamma c\bar{c}$ & $0.44\pm 0.26\pm 0.79$ &$0.11\pm 0.06\pm 0.17$\\ \hline
total             & $5.5\pm  1.5 \pm 1.6$  &$1.24\pm 0.35\pm 0.38$\\ 
\end{tabular}
\caption{Summary of the estimates of the background at 
high $M(b,j)$ in the boxes in the $M(\gamma,b,j)-M(b,j)$ plane 
defined in Figure \protect\ref{figscatmone}.}
\label{fivetot}
\end{table}

The result is that we expect $5.5\pm 1.5\pm 1.6$ events
in the big box, completely consistent with five observed.
We expect $1.2\pm 0.35\pm 0.38$ events in the small box.
The probability of observing five in the small box 
is 1.6\%, a $2.7\sigma$ effect, {\it a posteriori}.

\begin{figure}[ht]
\epsfxsize=8.6cm
\hspace*{3.5cm}
\epsfbox[23 152 536 672]{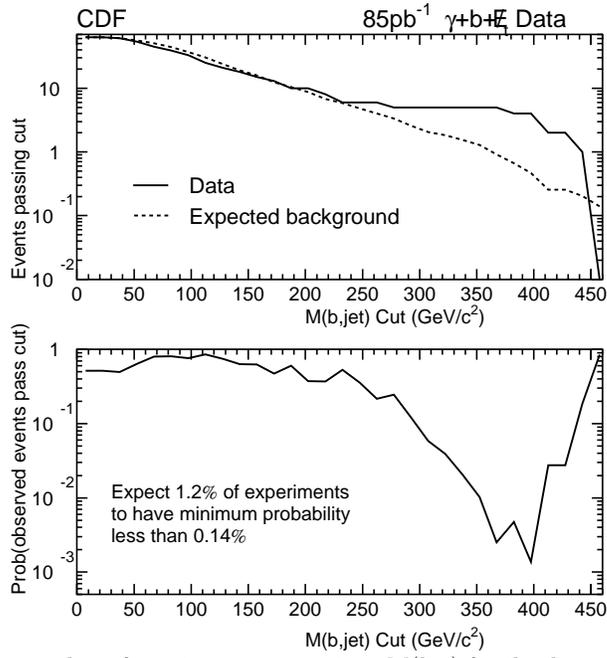}
\caption{The upper plot is the number of events passing 
a cut on $M(b,j)$ for the data and the positive tagging prediction.
The lower plot is the probability that the number of events passing
a cut on $M(b,j)$ is consistent with the positive tagging prediction.
The expected number of experiments with such a low minimum probability
is derived from 10,000 simulated experiments drawn from 
the distribution of the expected background.
}
\label{probfive}
\end{figure}

We next address a method for avoiding the bias in deciding where to
place a cut when estimating backgrounds to events on the tail of a
distribution.  This method was developed by the Zeus collaboration for
the analysis of the significance of the tail of the $Q^2$
distribution\cite{zeus}.  Figure \ref{probfive} summarizes this
method.  The Poisson probability that the background fluctuated to the
observed number of events (including uncertainties on the background
estimate) is plotted for different cut values.  We use the
projection of the scatter plot onto the $M(b,j)$ axis and make the cut
on this variable since this is where the effect is largest.  We find
the minimum probability is $1.4\times 10^{-3}$, which occurs for a cut
of $M(b,j)>$400~GeV.  We then perform 10,000 ``pseudo-experiments''
where we draw the data according to the background distribution
derived above and find the minimum probability each time.  We find
1.2\% of these experiments have a minimum probability lower than the
data, corresponding to a $2.7\sigma$ fluctuation.  Including the
effect of the uncertainties in the the background estimate does not
significantly change the answer.

We note that this method is one way of avoiding the bias from deciding
in what region to compare data and backgrounds after seeing the data
distributions.  It does not, however, remove the bias from the fact
that we are investigating this plot, over all others, because it looks
potentially inconsistent with the background.  If we make enough plots
one of them will have a noticeable fluctuation.  We conclude that the
five events on the tail represent something less than a $2.7\sigma$
effect.

\clearpage

\clearpage
\clearpage

\subsection{Additional Objects in the Data Sample} 
\label{section:otherobs}

\begin{figure}[ht]
\epsfxsize=8.6cm
\hspace*{3.5cm}
\epsfbox{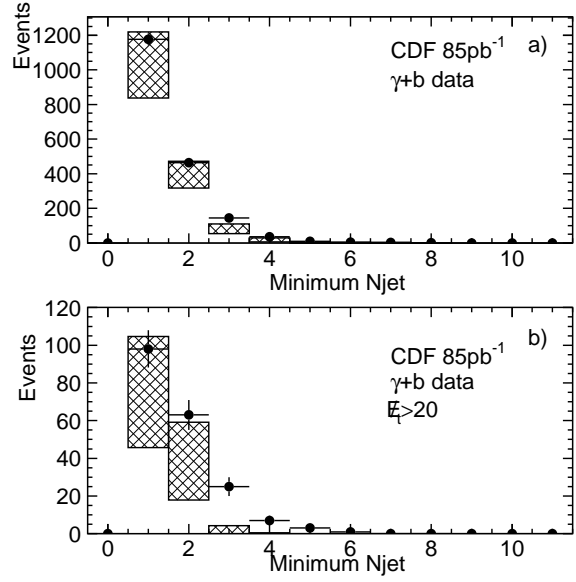}
\caption{The distribution of the number of events with 
$N$ or more jets, represented by the solid points.  The boxes are
centered on the background prediction and their size reflects
plus and minus one-$\sigma$ of combined statistical and systematic
uncertainty on the background prediction.  The distributions are: a)
all events with a photon and $b$-tagged jet, b) all events with a
photon, $b$-tagged jet and $\met>20$~GeV.  Some background predictions
are negative due to the large statistical fluctuations on the fake
photon background method.  The results are also tabulated in Table
\protect\ref{tab:njetandbg}.}
\label{fignjetplus}
\end{figure}

We have searched the $\gamma b$ data sample for other unusual
characteristics.  The creation and decay of heavy squarks, for
example, could produce an excess of events with multiple jets.  In
Figure \ref{fignjetplus} we histogram the number of events with $N$ or
more jets. Table \ref{tab:njetandbg} presents the numbers of events
observed and expected.  Some backgrounds are negative due to the large
statistical fluctuations of the fake photon background.  When all 
backgrounds are included  the distribution in the number of
jets in the data is consistent with that from background.

\begin{table}[ht]
\centering
\begin{tabular}{|c|c|c|c|c|}
Min $N_{jet}$ & Observed, $\met >0$~GeV  & Expected, $\met >0$~GeV  
	      & Observed, $\met >20$~GeV & Expected, $\met >20$~GeV   \\ \hline
1& 1175 & $1040\pm72\pm172$    & 98 & $77\pm23\pm20$ \\ \hline
2&  464 &  $394\pm44\pm63$     & 63 & $39\pm18\pm12$ \\ \hline
3&  144 &   $82\pm24\pm14$     & 25 & $-8\pm12\pm3$  \\ \hline
4&   36 &   $17\pm11\pm3$      &  7 &  $-$   \\ \hline
5&   10 &   $-$       &  3 & $-$ \\ \hline
6&    5 &   $-$  &  1 & $-$ \\ \hline
7&    2 &   $-$  &  0 & $-$            \\ \hline
8&    1 &   $-$  &  0 & $-$            \\
\end{tabular}
\caption{Numbers of events with $N$ or more jets 
and the expected Standard Model background.  Some background predictions
are negative due to the large statistical fluctuations on the 
fake photon background method.}
\label{tab:njetandbg}
\end{table}

We have searched in the sample of events with a photon and $b$--tagged jet
for additional high--$E_t$ objects using the requirements defined in 
Section \ref{section:otherselection}.
We find no events containing a second photon.
We find no events containing a hadronic $\tau$ decay or a muon.
We find one event with an electron; its characteristics are
listed in Table \ref{tab:electronevent}. 
In scanning this event, we note nothing else unusual about it.

We find 8 events of the 1175 which have a photon and $b$--tagged jet contain a
second $b$--tagged jet with $E_t>$30  GeV. (Out of the 1175, only 200
events have a second jet with $E_t>$30 GeV.)  Unfortunately, this is such a
small sample that we cannot use the background calculation to find the expected
number of these events (the photon background CES--CPR method returns 100\%
statistical  uncertainties).  One of the events with two tags has 30~GeV of
missing $E_t$ so is in the 98--event $\met>20$~GeV sample.

\begin{table}[ht]
\centering
\begin{tabular}{|c|c|c|c|c|c|c|c|}
Run/Event & $\gamma~E_t$ & \met & $M(\gamma,e)$ &b$~E_t$ & electron $E_t$ 
& $\Delta\phi(\gamma-\met)$  & $H_t$ \\ \hline
  63149/4148 &  42 & 17 & 21 & 106 & 33 & 43 & 212 \\ 
\end{tabular}
\caption{Characteristics of the one event with 
a photon, tagged jet, and an electron.}
\label{tab:electronevent}
\end{table}

\clearpage

\section{Limits on Models of Supersymmetry} 

In the following sections we present limits on three specific 
models of supersymmetry \cite{review}. 
Each of these models predicts significant numbers
of events with a photon, a $b$--quark jet and missing transverse
energy ({\it i.e} $\gamma b \met)$.

As is typical for supersymmetry models, each of these shows the
problems in the process of choosing a model and presenting limits on
it.  Each of these models is very specific and thus represents a very
small area in a very large parameter space. Consequently the odds that
any of these is the correct picture of nature is small.  They are
current theories devised to address current concerns and may appear
dated in the future. (This aspect is particularly relevant to the
experimentalists, who often publish their data simultaneously with an
analysis depending on a current model.)  In addition these models 
can show sensistivity to small changes in the parameters.

The first model is based on a particular location in MSSM  parameter 
space which 
produces the signature of $\gamma b \met +X$.   
We consider both direct production 
of charginos and neutralinos and, as a second model,  indirect production of
charginos and neutralinos  through squarks and gluinos. The third model 
is based on the gauge--mediated concept, discussed further below.

\subsection{
${{\tilde\chi}^0_2}\rightarrow\gamma {{\tilde\chi}^0_1}$ Model} 

This theoretical model was originally proposed in the context
of the anomalous CDF 
$ee\gamma\gamma\met$ event\cite{diphoton,gktop}. Here, however, 
we go beyond the constraints of this single event and only retain the 
essential elements of the model, optimized for CDF detector 
acceptence and efficiency. 
This is an MSSM model without 
any specific relation to a high--energy theory.
It does not assume high--energy constraints such as the unification of the 
sfermion or scalar masses as is assumed in the 
models inspired by Supergravity (SUGRA) \cite{review}.
In this section and the following section we develop
a baseline model point in parameter space.  The final limits
on this model will be found for this point and for some variations
around this point.

\subsubsection{Direct Gaugino Production in 
${{\tilde\chi}^0_2}\rightarrow\gamma {{\tilde\chi}^0_1}$ Model} 

The first part of the model\cite{gktop} 
is a light stop squark ($\tilde t$), the superpartner to the 
top quark.  In this model the ${\tilde \chi}^\pm_i$ then decays 
to $\tilde{t} b$ and the $\tilde t$ decays to ${\tilde\chi}^0_1 c$.
The second important feature of the model is the decay 
${{\tilde\chi}^0_2}\rightarrow \gamma {{\tilde\chi}^0_1}$ which 
dominates in a particular region of MSSM parameter space.
With these decays dominating, any event where a ${\tilde \chi}^\pm_i$ 
and a ${{\tilde\chi}^0_2}$ is produced, either directly or indirectly
through the strong production and decays of squarks and gluinos, 
will contain a photon, a $b$--quark jet and missing $E_t$.

The heart of the model\cite{gktop} is the decay 
${{\tilde\chi}^0_2}\rightarrow \gamma {{\tilde\chi}^0_1}$
so we examine in detail the parameter space where this decay dominates.
The branching ratio of this decay is large when one of the 
neutralinos is a pure photino and one is pure Higgsino.
To make a pure photino, we set $M_1=M_2$. The photino 
mass is then equal to $M_2$.
To make a pure Higgsino we set $\tan\beta\approx 1$.  To avoid the 
theoretical bias against
a very small $\tan\beta$ (which makes the top Yukawa coupling go to
infinity before the GUT scale) we will use $\tan\beta=1.2$.
In this case, the Higgsino mass is approximately equal to $|\mu|$.
The above is purely a result of the form of the neutralino mass 
matrix.  For definitions of these model parameters and discussions
of their roles in SUSY models please see~\cite{review}.

\begin{figure}[!ht]
\epsfxsize=8.6cm
\hspace*{3.5cm}
\epsfbox{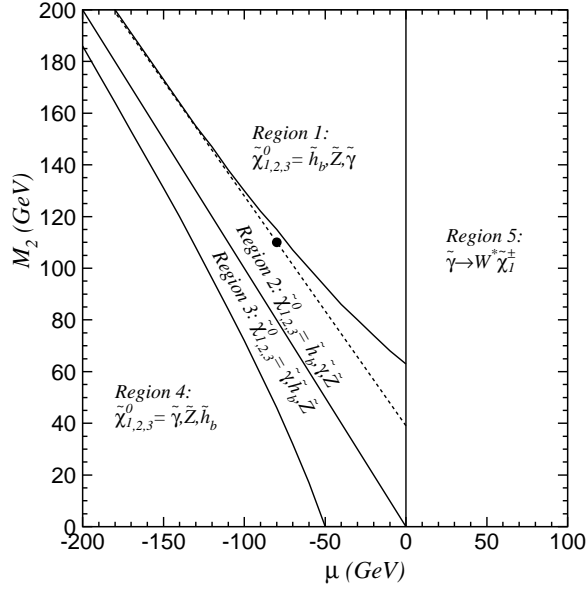}
\caption{The five regions in the supersymmetry parameter 
$M_2-\mu$ plane where different mass hierarchies occur.  The three
lightest neutralinos are denoted ${{\tilde \chi}}^0_{1,2,3}$,
respectively.  The dashed line is the locus of points scanned for the
limits and is given by $M_2=0.89|\mu|+39$~GeV.  The dot is the
baseline model described in the text.
}
\label{fig:mumtwo}
\end{figure}

This leaves two free parameters to define the charginos and
neutralinos, $M_2$ and $\mu$.  Figure \ref{fig:mumtwo} shows five
regions in the $\mu-M_2$ plane; Table \ref{tab:mumtwo} summarizes the
regions.  First we note that in region~5 ($\mu>0$) we do not observe
the decay ${{\tilde\chi}^0_2}\rightarrow \gamma {{\tilde\chi}^0_1}$
because typically ${{\tilde\chi}_1^\pm}<{{\tilde\chi}^0_2}$ and
${{\tilde\chi}^0_2}\rightarrow W^*{{\tilde\chi}_1^\pm}$.

For $\mu<0$, there are four regions.
In region 2, which is the region suggested in \cite{gktop},
the ${{\tilde\chi}^0_2}$ is the photino, ${{\tilde\chi}^0_1}$ 
is the Higgsino, and the decay 
${{\tilde\chi}^0_2}\rightarrow \gamma {{\tilde\chi}^0_1}$ dominates.
In region 3, ${{\tilde\chi}^0_2}$ is the Higgsino and 
${{\tilde\chi}^0_1}$ is the photino and 
the photon decay 
still dominates.  In region 1 the photino has become so heavy it is now 
the ${\tilde\chi}^0_3$.  In region 4, the Higgsino has become the 
${\tilde\chi}^0_3$.  In regions
3 and 4 it is still possible to get photon decays, sometimes even 
${\tilde\chi}^0_3\rightarrow \gamma {{\tilde\chi}^0_2}$.

\begin{table}[ht]
\centering
\begin{tabular}{|c|c|c|c|c|} 
Region & ${{\tilde\chi}^0_1}$ & ${{\tilde\chi}^0_2}$ & 
   ${\tilde\chi}^0_3$ & ${\tilde\chi}^0_4$ \\ \hline
1 & $\tilde h_b$ & $\tilde Z$ & $\tilde \gamma$ & $\tilde h_a$ \\ \hline
2 & $\tilde h_b$ & $\tilde \gamma$ &  $\tilde Z$ & $\tilde h_a$ \\ \hline
3 & $\tilde \gamma$ &  $\tilde h_b$ & $\tilde Z$ &  $\tilde h_a$ \\ \hline
4 & $\tilde \gamma$ &  $\tilde Z$ &  $\tilde h_b$ &  $\tilde h_a$ \\ 
\end{tabular}
\caption{The approximate content of the neutralinos in the four regions
of the $\mu-M_2$ plane with $\mu < 0$ 
shown in Figure \protect\ref{fig:mumtwo}. 
The symbols $\tilde h_a$ and $\tilde h_b$ 
are simply the antisymmetric and symmetric 
combinations of $\tilde H^0_1$ and $\tilde H^0_2$\protect\cite{review}.}
\label{tab:mumtwo}
\end{table}

We choose to concentrate on region 2 where the photon plus $b$ 
decay signature 
can be reliably estimated by the Monte Carlo 
event generator {\tt PYTHIA} \cite{Pythia}.  The 
${{\tilde\chi}^0_2}\rightarrow \gamma {{\tilde\chi}^0_1}$ decay dominates here.
We also note that in this region
the cross section for ${{\tilde\chi}_2^\pm}{{\tilde\chi}^0_2}$ is 
3--10 times 
larger than the cross section for ${{\tilde\chi}_1^\pm}{{\tilde\chi}^0_2}$ 
even though the
${{\tilde\chi}_2^\pm}$ is significantly heavier than the 
${{\tilde\chi}_1^\pm}$.  This is due to the 
large $\tilde W$ component of the ${{\tilde\chi}_2^\pm}$.

Since region 2 is approximately one-dimensional, we scan in only
one dimension, along the diagonal, when setting limits on 
${{\tilde\chi}_2^\pm}{{\tilde\chi}^0_2}$ production.  To decide where 
in the region 
to place the model, we note that the mass of ${{\tilde\chi}^0_2}$ equals 
$M_2$ and the mass of 
${{\tilde\chi}^0_1}=|\mu|$ in this region.  To give the photon added boost 
for a greater sensitivity, we will set $M_2$ significantly larger
than $|\mu|$.  This restricts us to the upper part of region 2.
The dotted line in Figure \ref{fig:mumtwo} is the set of points defined
by all these criteria and is given by $M_2=0.89*|\mu|+39$~GeV.

The next step is to choose a $\tilde t$ mass.  It is necessary that
${{\tilde\chi}^0_1}<\tilde t < {{\tilde\chi}_1^\pm}$ for the decay
${{\tilde\chi}_1^\pm},{{\tilde\chi}_2^\pm}\rightarrow b \tilde t$ to
dominate.  We find that in Region 2, ${{\tilde\chi}_1^\pm}\approx
M_2$.  If the $\tilde t$ mass is near the ${{\tilde\chi}_1^\pm}$, the
$b$ will only have a small boost, but the ${{\tilde\chi}^0_1}$ in the
decay $\tilde t \rightarrow c {{\tilde\chi}^0_1}$ will have a greater
boost, giving greater $\met$.  If the $\tilde t$ mass is near the
${{\tilde\chi}^0_1}$, the opposite occurs.  In Monte Carlo studies, we
find considerably more sensitivity if the $\tilde t$ mass is near the
${{\tilde\chi}^0_1}$.  We set the $\tilde t$ mass to be
$M_{{{\tilde\chi}^0_1}}+5$~GeV.  Since the
${{\tilde\chi}_2^\pm}{{\tilde\chi}^0_2}$ production cross section is
larger than ${{\tilde\chi}_1^\pm}{{\tilde\chi}^0_2}$ and will be
detected with better efficiency, when we simulate direct production we
set the Monte Carlo program to produce only
${{\tilde\chi}_2^\pm}{{\tilde\chi}^0_2}$ pairs.  The final limit is
expressed as a cross section limit plotted versus the
${{\tilde\chi}_2^\pm}$ mass (which is very similar to the
${{\tilde\chi}^0_2}$ mass).  This model is designed to provide a
simple, intuitive signature that is not complicated by branching
ratios and many modes of production.

For the baseline model, we chose a value of $\mu$ near the exclusion 
boundary of current limits \cite{stoplimit} on a $\tilde t$ which
decays to $c\chi^0_1$.  The point we chose is 
$M_{{{\tilde\chi}^0_1}}=80$~GeV.  
From the above prescription, this corresponds to
$M_{{{\tilde\chi}^0_1}}=-\mu=80$~GeV,
$M_{{{\tilde\chi}^0_2}}=M_{{{\tilde\chi}_1^\pm}}=M_2=110$~GeV, and
$M_{\tilde t}=85$~GeV.  This point, indicated by the dot in Figure
\ref{fig:mumtwo}, gives the lightest mass spectrum with good mass
splittings that is also near the exclusion boundary from LEP and DO
Collaborations.

\subsubsection{Squarks and Gluinos}   
\label{section:sqandgl}

Now we address the squarks and gluinos, which can produce 
${{\tilde\chi}_i^\pm}{{\tilde\chi}^0_2}$ in their decays, and sleptons, 
which can appear in the decays of charginos and neutralinos.  

We will set the squarks  (the lighter $\tilde b$ and both left and right 
$\tilde u$, $\tilde d$, $\tilde s$ and $\tilde c$) to 200~GeV and the gluino
to 210~GeV.  The heavier $\tilde t$ and $\tilde b$ are above 1 TeV.
The gluino will decay to the squarks and their respective quarks.
The squarks will decay to charginos or neutralinos 
and jets.  This will maximize
the production of ${{\tilde\chi}_i^\pm}{{\tilde\chi}^0_2}$ and therefore 
the sensitivity.

This brings us to the limit on indirect production in the 
${{\tilde\chi}^0_2}\rightarrow\gamma{{\tilde\chi}^0_1}$ model.  
The chargino and neutralino parameters are fixed 
at the baseline model parameters.  We then vary 
the gluino mass and set the squark mass according to 
$M_{\tilde g}=M_{\tilde Q}+10$~GeV.
The limit is presented as a limit on cross section 
plotted versus the gluino mass.  
When the gluino mass crosses the 
$\tilde t\bar t$ threshold at 260~GeV, the gluino can decay to 
$\tilde t\bar t$ and 
production of 
${{\tilde\chi}_i^\pm}{{\tilde\chi}^0_2}$ decreases.
However, since all squarks are lighter than the gluino, the 
branching ratio to the $\tilde t$ is limited and 
production will not fall dramatically.

Some remaining parameters of the model are now addressed.
Sleptons could play a role in this model. 
They have small cross sections so they are not often directly produced, 
but if the sleptons are lighter than the charginos, the charginos 
can decay into the sleptons.  In particular, the chargino
decay $\tilde t b$ may be strongly suppressed if it competes with 
a slepton decay. 
We therefore set the sleptons to be very heavy so they do not compete for 
branching ratios.  We set $M_A$ large.  
The lightest Higgs turns out to be only 
87~GeV due to the corrections from the light third--generation squarks.
This is below current limits so we attempted to tune 
the mass to be heavier and found it was difficult to achieve, 
given the light $\tilde t$ and low $\tan\beta$.

Using the {\tt PYTHIA } 
Monte Carlo program, we find that 69\% of 
all events generated with squarks and gluinos
have the decay 
${{\tilde\chi}^0_2}\rightarrow \gamma {{\tilde\chi}^0_1}$,
58\% have the decay ${{\tilde\chi}_i^\pm}\rightarrow {\tilde t} b$, and 
30\% have both. (To be precise, the light stop squark was excluded from
this exercise, as it decays only to $c \chi^0_1$. A light stop pair thus
gives the signature $\ccbar+\met$, one of the signatures used to 
search for it,\cite{stoplimit,ccmet} but not of interest here.)

\clearpage
\subsubsection{Acceptances and Efficiencies}   
\label{sec:eff}

This section describes the evaluation of the 
acceptance and efficiency for the indirect production
of ${{\tilde\chi}_i^\pm}{{\tilde\chi}^0_i}$ through squarks and gluinos 
and the direct production of ${{\tilde\chi}_2^\pm}{{\tilde\chi}^0_2}$ 
in the MSSM model of ${{\tilde\chi}^0_2}\rightarrow \gamma {{\tilde\chi}^0_1}$.
We use the {\tt PYTHIA} Monte Carlo with the CTEQ4L parton
distribution functions (PDFs)\cite{CTEQ4L}.
The efficiencies for squark and gluino production 
at the baseline point are listed in Table \ref{tab:eff}.

\begin{table}[ht]
\centering
\begin{tabular}{|c|c|c|} 
Cuts & &Cumulative Efficiency (\%) \\ \hline
Photon & $E_t>25$~GeV,$|\eta|<1.0$, ID cuts & 50 \\
One jet   & $E_{t,corr}>30$~GeV, $|\eta|<2.0$ & 47 \\
One SVX tag  & $E_{t,corr}>30$~GeV, $|\eta|<2.0$ & 4.3 \\
$\met$ & $> 40$~GeV& 2.9 \\ 
\end{tabular}
\caption{Efficiencies for the baseline point with squark and gluino 
production.  The efficiencies do not include branching ratios.}
\label{tab:eff}
\end{table}

The total efficiencies, which will be used to set production limits below, are
listed in  Table \ref{tab:sg} for the production of 
${{\tilde\chi}_i^\pm}{{\tilde\chi}^0_i}$ through squarks and gluinos,  and in
Table \ref{tab:cn} for direction production. Typical efficiencies are 2-3\%
in the former case, and 1\% in the latter.

\subsubsection{Systematic Uncertainty}   
\label{sec:syst}

Some systematics are common to the indirect production and the 
direct production.
The efficiencies of the isolation requirement in the Monte Carlo and
$Z\rightarrow e^+e^-$ control sample cannot be
compared directly due to differences in the $E_t$-spectra of the 
electromagnetic
cluster, and the multiplicity and $E_t$ spectra of associated jets.
The difference (14\%) is taken to be the
uncertainty in the efficiency of the photon identification cuts.
The systematic uncertainty on the $b$--tagging efficiency (9\%) is
the statistical uncertainty in comparisons of the Monte Carlo and data.
The systematic uncertainty on the luminosity (4\%) reflects the
stability of luminosity measurements.

We next evaluate systematics specifically for the indirect
production. The baseline parton distribution function is CTEQ4L.
Comparing the efficiency with this PDF to the efficiencies obtained
with MRSD0$^\prime$~\cite{MRSD} and GRV--94LO~\cite{GRV} for the
squark and gluino production, we find a standard deviation of
5\%. Turning off initial-- and final--state radiation (ISR/FSR) in the
Monte Carlo increases the efficiency by 1\% and 2\% respectively and
we take half of these as the respective systematics. Varying the jet
energy scale by 10\% causes the efficiency to change by 4\%. In
quadrature, the total systematic for the indirect production is 18\%.

Evaluating the same systematics for the direct production, we find 
the uncertainty from the choice of PDF 
is 5\%, from ISR/FSR is 2\%/9\%, and from jet energy scale is 4\%.
In quadrature, the total systematic uncertainty 
for the direct production is 20\%.

\subsubsection{Limits on the 
${{\tilde\chi}^0_2}\rightarrow\gamma{{\tilde\chi}^0_1}$ Model, 
Indirect Production}  

To calculate an approximate upper limit on the number of $\gamma b
\met$ events from squark and gluino production, we use the limit
implied from the observed 2 events, including the effect of the 
systematic uncertainties\cite{Helene,Zeck}. We
divide the Poisson probability for observing $\leq$2 events for a
given expected signal and background, convoluted with the
uncertainties, by the Poisson probability for observing $\leq$2
events for a given background only, also convoluted with the
uncertainties.  The number of expected signal events is increased
until the ratio falls below 5\%, leading to an approximate 95\%
confidence level limit of 6.3 events.  Other limits in this paper are
computed similarly.

\begin{figure}[!ht]
\epsfxsize=8.6cm
\hspace*{3.5cm}
\epsfbox{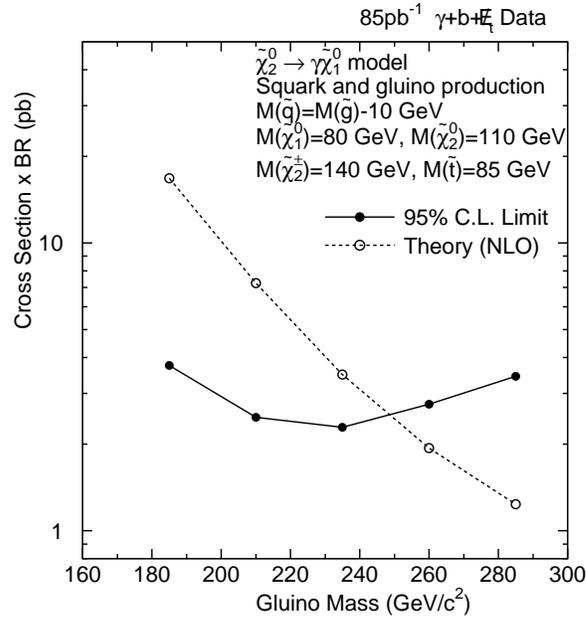}
\caption{The limits on the cross section times branching ratio
 for SUSY production of $\gamma b\met$ events in the
 ${\tilde\chi}^0_2\rightarrow\gamma{\tilde\chi}^0_1$ model. 
All production processes
 have been included; the dominant mode is the production of squarks
 and gluinos which decay to charginos and neutralinos.  The overall
 branching ratio to the $\gamma b \met$ topology is approximately
 30\%.}
\label{fig:sglimit}
\end{figure}

This upper limit, the efficiency described above (also see Table
\ref{tab:sg}), and the luminosity, 85~pb$^{-1}$, are combined to find
the cross section limit for this model.  The theoretical cross section
is calculated at NLO using the {\tt PROSPINO} program \cite{prospino}.
The effect is to uniformly increase the strong interaction production
cross sections by 30\% (improving the limit).  At the baseline point
(including squarks and gluinos) described above, we expect 18.5
events, so this point is excluded.  Next we find the limit as a
function of the gluino mass.  The squark mass is 10~GeV below the
gluino mass and the rest of the sparticles are as in the baseline
point.  We can exclude gluinos out to a mass of 245~GeV in this model.
The limits are displayed in Table
\ref{tab:sg} and Figure \ref{fig:sglimit}.

\begin{table}[ht]
\centering
\begin{tabular}{|c|c|c|c|c|} 
$M_{\tilde g} (GeV) $  & $M_{\tilde q}$ (GeV) ,  & 
    $\sigma_{th}\times BR$ (pb) & $A\epsilon~(\%)$ & 
$\sigma_{95\%~lim} \times BR$ (pb) \\ \hline
185  & 175 & 16.8 & 1.97 & 3.76  \\ \hline
210  & 200 & 7.25 & 2.98 & 2.49  \\ \hline
235  & 225 & 3.49 & 3.23 & 2.30  \\ \hline
260  & 250 & 1.94 & 2.69 & 2.76 \\ \hline
285  & 275 & 1.24 & 2.16 & 3.45  \\ 
\end{tabular}
\caption{Efficiency times acceptance and limits on 
indirect production of ${{\tilde\chi}_i^\pm}{{\tilde\chi}^0_2}$ though
squarks and gluinos.  Approximately 30\% of events contain the decays
${{\tilde\chi}^0_2}\rightarrow \gamma {{\tilde\chi}^0_1}$ and
${{\tilde\chi}_i^\pm}\rightarrow \tilde t b$.  The efficiencies in
this table are found as the number of events passing all cuts divided
by the number of events that contain both of these decays.  The
product of cross section times branching ratio listed in each case is
for all open channels of SUSY production.  Masses are given in GeV 
(following our convention of quoting $M \times c^2$) and 
cross sections are in pb.  The second row is the baseline point.}
\label{tab:sg}
\end{table}

\subsubsection{Limits on the 
${{\tilde\chi}^0_2}\rightarrow\gamma{{\tilde\chi}^0_1}$ Model, 
Direct Production}  

\begin{figure}[!ht]
\epsfxsize=8.6cm
\hspace*{3.5cm}
\epsfbox{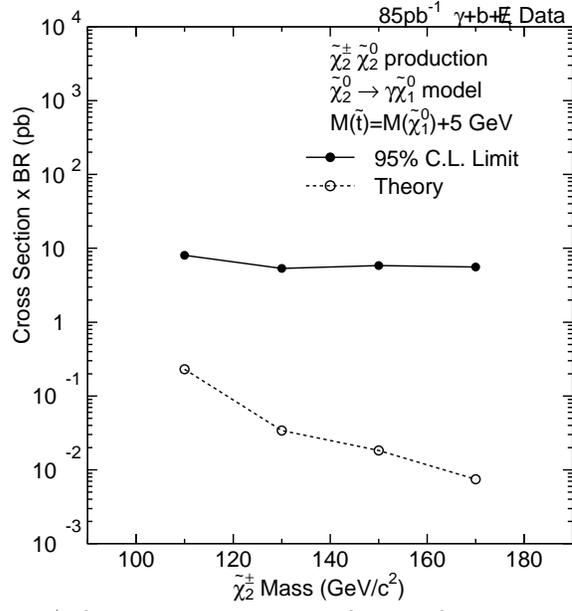}
\caption{The limits on the ${{\tilde\chi}_2^\pm}{{\tilde\chi}^0_2}$ 
cross section in the ${\tilde\chi}^0_2\rightarrow\gamma{\tilde\chi}^0_1$ 
SUSY
model.  The branching ratios ${{\tilde\chi}^0_2}\rightarrow\gamma
{{\tilde\chi}^0_1}$ and ${\tilde \chi}^\pm_i\rightarrow \tilde{t} b
\rightarrow ({\tilde\chi}^0_1 c)b$ are taken to be 100\%.}
\label{fig:cnlimit}
\end{figure}

In this case the number of observed events (two) is
convoluted with the systematic 
uncertainty to obtain an upper limit of 6.4 events (95\% C.L.).
To calculate the expected number of events from the direct production of 
${{\tilde\chi}_2^\pm}{{\tilde\chi}^0_2}$, 
we vary $\mu$, and calculate the $M_{\tilde t}$ and $M_2$ 
as prescribed above.  The results are shown in 
Table~\ref{tab:cn} and Figure~\ref{fig:cnlimit}.
For these values of the model parameters, the branching ratios
${{\tilde\chi}_2^\pm}\rightarrow \tilde{t} b$ and
${{\tilde\chi}^0_2}\rightarrow\gamma {{\tilde\chi}^0_1}$ are 100\%.

As can be seen from Figure~\ref{fig:cnlimit}, the predicted rates from 
direct production are smaller than the measured limits by 1-3 
orders-of-magnitude, and no mass limit on the ${{\tilde\chi}_2^\pm}$ mass can
be set.

\begin{table}[ht]
\centering
\begin{tabular}{|c|c|c|c|c|c|c|c|} 
$M_{{{\tilde\chi}^0_1}}=-\mu$  & $M_{{{\tilde\chi}^0_2}}=M_2$,  
& $M_{{{\tilde\chi}_1^\pm}}$ & $M_{{{\tilde\chi}_2^\pm}}$ & $M_{\tilde t}$ & 
    $\sigma_{th}$ & $A\epsilon~(\%)$ & $\sigma_{95\%~lim}$ \\ \hline
25  &  61 &  71 & 110 &  30 & 0.23   & 0.93  & 8.06  \\ \hline
62  &  95 &  94 & 130 &  67 & 0.034  & 1.41  & 5.33  \\ \hline
79  & 110 & 108 & 140 &  85 & 0.018  & 1.29  & 5.85  \\ \hline
93  & 123 & 118 & 150 &  98 & 0.0075 & 1.34  & 5.58  \\ \hline
118 & 146 & 140 & 170 & 123 & 0.0022 & 1.27  & 5.94  \\ 
\end{tabular}
\caption{Efficiencies and limits on 
direct production of ${{\tilde\chi}_2^\pm}{{\tilde\chi}^0_2}$.  
Branching ratios ${{\tilde\chi}^0_2}\rightarrow\gamma {{\tilde\chi}^0_1}$ and 
${\tilde \chi}^\pm_i\rightarrow \tilde{t} b \rightarrow ({\tilde\chi}^0_1 c)b$
are 100\%.
Masses are in GeV and the cross sections are in pb.  
The third row is the baseline point.}
\label{tab:cn}
\end{table}

\clearpage

\clearpage

\clearpage

\clearpage

\subsection{Gauge-mediated Model} 

This is the second SUSY model\cite{review} which can give substantial
production of the signature $\gamma b \met$.  In this model the
difference between the mass of the standard model particles and their
SUSY partners is mediated by gauge (the usual electromagnetic, 
weak, and strong)
interactions \cite{gauge} instead of gravitational interactions as in
SUGRA models.  SUSY is assumed broken in a hidden sector.  Messenger
particles gain mass through renormalization loop diagrams which
include the hidden sector.  SUSY particles gain their masses through
loops which include the messenger particles.

This concept has the consequence that the strongly--interacting
squarks and gluinos are heavy and the right-handed sleptons are at the
same mass scale as the lighter gauginos.  A second major consequence
is that the gravitino is very light (eV scale) and becomes the LSP.
The source of $b$ quarks is no longer the third generation squarks,
but the decays of the lightest Higgs boson.  If the lightest
neutralino is mostly Higgsino, the decay
${{\tilde\chi}^0_1}\rightarrow h \tilde G$ can compete with the decays
${{\tilde\chi}^0_1}\rightarrow Z \tilde G$ and
${{\tilde\chi}^0_1}\rightarrow \gamma \tilde G$.  The Higgs decays to
$b\bar{b}$ as usual.  Since SUSY particles are produced in pairs, each
event will contain two cascades of decays down to two
${{\tilde\chi}^0_1}$'s, each of which in turn will decay by one of these
modes.  If one decays to a Higgs and one decays to a photon, the event
will have the signature of a photon, at least one $b$--quark jet,
and missing $E_t$.

We will use a minimal gauge-mediated model with one exception.
This MGMSB model has five parameters, with the following values:
\begin{itemize}
\item $\Lambda = 61-90$~TeV, the effective SUSY-breaking scale;
\item $M/\Lambda=3$, where $M$ is the messenger scale;
\item $N=2$ the number of messenger multiplets;
\item $\tan\beta=3$;
\item the sign of $\mu<0$.
\end{itemize}
We will compute the MGMSB model using the 
GMSB option of {\tt ISAJET} \cite{isajet}.
We then re--enter the model using the MSSM options so that we can 
make one change:
we set \mbox{$\mu=-0.75M_1$}.  This makes the lightest neutralino a Higgsino
so the branching ratio for 
${{\tilde\chi}^0_1}\rightarrow h \tilde G$ will be competitive
with ${{\tilde\chi}^0_1}\rightarrow \gamma \tilde G$.
We produce all combinations of ${{\tilde\chi}_i^\pm}$ and 
${{\tilde\chi}^0_j}$, which are the only significant
cross sections.
We vary $\Lambda$ which varies the overall mass scale of the 
supersymmetric particles.

The model masses and
branching ratios are given in Table \ref{tab:gmsbeff}.  
The branching ratio is defined as the number of events with 
${{\tilde\chi}^0_1}\rightarrow \gamma \tilde G$ and 
${{\tilde\chi}^0_1}\rightarrow h \tilde G$ divided by the number of 
events produced from all sources predicted by the model.
We are using {\tt ISAJET} with the  CTEQ4L parton distribution function.  
The first point appears to have an unusually large efficiency 
because of other sources for $b$ quarks 
which are not reflected in the definition of the 
signal branching ratio.
We use the 
systematic uncertainties evaluated using the direct production of the 
${{\tilde\chi}^0_2}\rightarrow\gamma {{\tilde\chi}^0_1}$ model.

\begin{figure}[!ht]
\epsfxsize=8.6cm
\hspace*{3.5cm}
\epsfbox{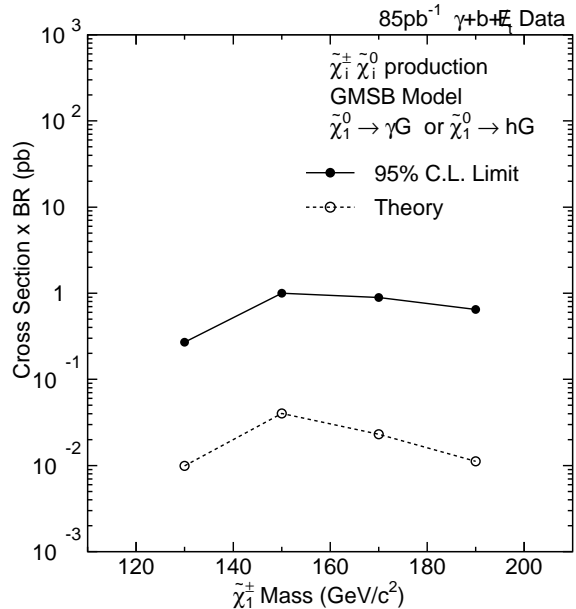}
\caption{The limits on the cross section times branching ratio
 for SUSY production of $\gamma b\met$ events in the GMSB model. All
 production processes have been included.}
\label{fig:gmsblimit}
\end{figure}

Taking the two events observed, and convoluting with a 20\% systematic
uncertainty gives an upper limit of 6.4 events observed at 95\% C.L.
The final limits on this model are presented in Table \ref{tab:gmsb}
and displayed in Figure~\ref{fig:gmsblimit}.  Again, one can see that
the experimental sensitivity is not adequate to set a mass limit (this time
on the ${{\tilde\chi}^{\pm}_1}$ mass) by several orders of magnitude.

\clearpage 

\begin{table}[ht]
\centering
\begin{tabular}{|c|c|c|c|c|} 
 $M_{{{\tilde\chi}^0_1}}$,  & $M_{{{\tilde\chi}^0_2}}$ 
& $M_{{{\tilde\chi}_1^\pm}}$ 
   & $BR({{\tilde\chi}^0_1}\rightarrow \gamma \tilde G)$  & 
$BR({{\tilde\chi}^0_1}\rightarrow h \tilde G)$ \\ \hline
 113 & 141 & 130 & 90 & 2   \\ \hline
 132 & 157 & 147 & 62 & 18  \\ \hline
 156 & 178 & 170 & 33 & 40  \\ \hline
 174 & 194 & 186 & 22 & 50  \\ 
\end{tabular}
\caption{The models used in the limits on the GMSB scenario.
The lightest Higgs boson is 100~GeV.   
The masses are in GeV and the branching ratios are in \%.}
\label{tab:gmsbeff}
\end{table}

\begin{table}[ht]
\centering
\begin{tabular}{|c|c|c|c|} 
 $A\epsilon$ & BR 
        & $\sigma_{th}\times BR$ & $\sigma_{95\%~lim}\times BR$ \\ \hline
  27.4 &  3  & 0.010   &  0.27  \\ \hline
  7.5 & 20  & 0.0402  &  1.00  \\ \hline
  8.4 & 23  & 0.0230  &  0.89  \\ \hline
 11.4 & 18  & 0.0111  &  0.66  \\ 
\end{tabular}
\caption{Efficiencies and limits on 
direct production of ${{\tilde\chi}_i^\pm}{{\tilde\chi}^0_j}$ 
in the GMSB scenario.
Branching ratios are not included in these efficiencies.
The first row has an inflated efficiency due to the definition 
of the branching ratio.
The units of $A\epsilon$ and the branching ratio are \% and the 
cross sections are in pb.}
\label{tab:gmsb}
\end{table}

\clearpage

\section{Model--independent Limits} 

As described in the introduction, there are several advantages to
presenting limits of searches in a model--independent form.  In the
previous sections we derived limits on models of supersymmetry and
presented the results as a limit on a cross section times branching
ratio for a specific model, $(\sigma\times BR)^{lim}=N^{lim}/{\cal
L}A\epsilon$, where $N_{lim}$ is the 95\% confidence level on the
number of events of anomalous production and ${\cal L}$ is the
integrated luminosity.  We make a distinction between the
acceptance, $A$, which is defined as probability that an object passes
$E_t$, $\eta$, and $\Delta\phi$ cuts, and the efficiency, $\epsilon$,
which is the probability of events surviving all other sources of
inefficiencies, such as photon identification cuts or $b$--tagging
requirements, which is detector-specific.  The acceptance may be
calculated from kinematic and geometric criteria alone, so an
experienced worker in the field can compute it using only a Monte Carlo
event generator program, while the efficiency requires access to the
full detector simulation and, typically, multiple control samples.  In
our formulation, $N_{lim}$ includes the degradation in sensitivity due
to uncertainties on $A\epsilon$, luminosity, and background
subtractions, when they are included, as well as the statistical upper
limit on the number of events.

In the case 
of model--independent limits, there is no model to determine the 
efficiency and therefore we report a limit on 
$(\sigma\times BR\times A\epsilon)^{lim}=N^{lim}/{\cal L}$.
These limits, which are presented in the next section, do not have an
immediate interpretation.  (They do imply, however, a cross section range that
we are {\it not} sensitive to, even with perfect efficiency.)
In order to determine the meaning of these limits,
in particular if a model is excluded or not, 
there must still be a mechanism for an interested physicist to 
calculate $A\epsilon$ for the model, 
and we develop three methods in the Appendix.

\subsection{Model--independent Limits on $\gamma b X$
Signatures} 

The limit on $(\sigma\times BR\times A \epsilon)^{lim}_i=N^{lim}/{\cal
L}$ is described by reporting that two events are observed with an
isolated photon with $E_t>$25~GeV and $|\eta|<1$, a SVX-tagged
$b$--quark jet with $E_t>$30~GeV and $|\eta|<2$,
$\Delta\phi(\gamma-\met)<168^\circ$ and $\met>40$~GeV.  The cuts are
fully described in Section \ref{section:dataselection}.  The 
integrated luminosity for this sample is ($85\pm 3$)~pb$^{-1}$.

The resulting 95\% confidence level 
limit on $(\sigma\times BR\times A\epsilon)^{lim}$
for the $\gamma b \met$ signature is then 0.069~pb.  
Adding the 4\% luminosity uncertainty 
we find the cross section limit increases to 0.070~pb.  
If we also add the 22\% uncertainty in $A\epsilon$ 
from the $WW$ limits (a typical uncertainty on an efficiency for
this signature) discussed in the Appendix,
we find the cross section limit increases 10\% to 0.077~pb. This is the
final model--independent limit on the signature $\gamma b \met$.
The limit on the $\gamma b$ signature before any $\met$ 
requirement is 5.9~pb and
the limit from the $\gamma b \met$ signature from the 98--event sample 
with $\met>20$~GeV is 0.99~pb.

The search for other objects in these events is described in 
Sections \ref{section:otherselection} and 
\ref{section:otherobs}.  When we find no events, we can set
a 95\% confidence level limit on $(\sigma\times BR\times \epsilon A)^{lim}$
of 0.038~pb assuming 4\% uncertainty in the luminosity
and 22\% uncertainty in the efficiency.  This would apply to the searches
for events with an additional photon, a muon or tau.
For events with an additional electron, we observe one event
and our limit becomes 0.057~pb.

For events with $N$ or more jets
as shown in Figure \ref{fignjetplus},
we find the limits listed
in Table \ref{tab:jetlim}.

\begin{table}[ht]
\centering
\begin{tabular}{|c|c|c|} 
$N$ & $\sigma_{lim}^{95\%}$ (pb), $\met>0$~GeV & 
$\sigma_{lim}^{95\%}$ (pb), $\met>20$~GeV \\ \hline
 1  &  5.9   & 0.99 \\ \hline
 2  &  2.5   & 0.77 \\ \hline
 3  &  1.5   & 0.50 \\ \hline
 4  &  0.52  & 0.18 \\ \hline
 5  &  0.24  & 0.10 \\ \hline
 6  &  0.14  & 0.062 \\ \hline
 7  &  0.083 & 0.038 \\ \hline
 8  &  0.062 & 0.038 \\ 
\end{tabular}
\caption{The 95\% confidence level limits on 
$(\sigma\times BR\times \epsilon A)^{lim}$
in pb for events with $N$ or more jets, including the $b$ jet.}
\label{tab:jetlim}
\end{table}

\section{Conclusions}

We have searched in 85~pb$^{-1}$ of CDF data for anomalous production
of events with a high--$E_t$ photon and a $b$--tagged jet. We find 1175
events with a photon with $E_t > 25$ GeV and a $b$--tagged jet with
$E_t > 30$ GeV, versus 1040$\pm$186 expected from standard model
backgrounds. Further requiring missing transverse energy $\met > 40$
GeV, in a direction not back-to-back with the photon ($\Delta\phi <
168^{\circ}$), we observed 2 events versus 2.3$\pm$ 1.6 events
expected.  In addition we search in subsamples of these events for
electrons, muons, and tau-leptons, additional photons, jets and
$b$--quark jets.  We conclude that the data are consistent with
standard model expectations.

We present limits on three current models of supersymmetry. The first
is indirect production of chargino-neutralino pairs through squark and
gluino production, where the photon is produced in $\kaitwo \goes
\gamma \kaione$ and the $b$-quark comes from the chargino decay into
the light stop squark $\kaioneplus \goes \stop b$. A choice of favorable
values of the parameters allows setting a lower mass limit on the
gluino mass of 250 GeV. The second model is similar, but we look only
at direct production of the $\kaitwo\chi^{\pm}_2$ pair. A cross
section limit of $\sim$ 7-10 pb is set, but is above the predictions
for all $\kaitwo$ masses so that no mass limit can be set. Lastly, a
GMSB model is considered in which the photon comes from the decay
$\kaione \goes \gamma\Gravitino$.  Limits in the range 0.3-1.0 pb are
set versus the mass of the $\kaione$, but again no mass limit can be
set as the cross section predictions are lower than the limit.

Finally, we present a model--independent limit of 0.077 pb on the
production of events containing the signature $\gamma b \met$, and
propose new methods for applying model--independent limits to models
that predict similar broad signatures.  We conclude that an experienced
model--builder can evaluate whether model-independent limits apply to
a particular model with an uncertainty of approximately 30\%.

\section{Acknowledgments}

     We thank the Fermilab staff and the technical staffs of the
participating institutions for their vital contributions.  This work was
supported by the U.S. Department of Energy and National Science Foundation;
the Italian Istituto Nazionale di Fisica Nucleare; the Ministry of Education,
Science, Sports and Culture of Japan; the Natural Sciences and Engineering 
Research Council of Canada; the National Science Council of the Republic of 
China; the Swiss National Science Foundation; the A. P. Sloan Foundation; the
Bundesministerium fuer Bildung und Forschung, Germany; the Korea Science 
and Engineering Foundation (KoSEF); the Korea Research Foundation; and the 
Comision Interministerial de Ciencia y Tecnologia, Spain.

\clearpage

\section{Appendix: Application of Model--independent Limits} 

In the body of this paper, we present the limits 
on specific models of new physics
that predict the $\gamma b \met$ signature, then rigorously 
calculate $A\epsilon$
for that model by using a Monte Carlo with a full detector simulation.
We present our limits on 
$(\sigma\times BR)^{lim}=N^{lim}/{\cal L}A\epsilon$, 
or a parameter of the model such 
as the mass of a supersymmetric particle.

A new paradigm, the signature--based 
or, equivalently, model--independent search may be 
an effective method for reporting the results of searches
in the future.
In this case, a signature, such as the photon and $b$--quark
jet addressed in this paper, is the focus of the 
search rather than the predictions of a particular model.

There are several advantages to this approach\cite{diphoton,cargese,sleuth}.  
\begin{enumerate}
\item The results are not dated by our current theoretical understanding.
\item No {\it a priori} judgement is necessary to
determine what is an interesting model.
\item The results more closely represent the experimental observations
and results will be presented in a form that
can be applied to a broad range of models including those not 
yet imagined.
\item The number of signatures is more reasonably limited than 
the number of models and model parameters.
\item Concentrating on a particular model can tend to 
focus the search very narrowly, ignoring variations 
on the signature which may be just as likely to yield a discovery.
\item Time spent on studying models can be diverted to systematically
searching additional signatures.
\end{enumerate}

In order to reflect the data results more generally, in the body
of this paper we also present a limit on 
$(\sigma\times BR \times A\epsilon)^{lim}=N^{lim}/{\cal L}$
for the signatures with no calculation of $A\epsilon$.  
With limits presented this way, the collaboration itself, model--builders
and other interested workers are no longer given limits on the physics
models directly but now must derive the limits themselves.
This has the potential for a wider application of the limits.
In a pratical sense, it means the interested workers 
must calculate $A\epsilon$ for the model under study.

In this Appendix, we present three methods to calculate 
$A\epsilon$.  These results together with the model-independent limits
can be used to set limits on most models 
that predict events with the $\gamma b \met$ signature.

The three methods are referred to as ``object efficiencies'', 
``standard model calibration process,'' and ``public Monte Carlo''.
In the sections below we describe each in turn.  In the following 
sections, we test these methods by comparing the results 
of each $A\epsilon$ calculation to the rigorously--derived $A\epsilon$ 
for the specific supersymmetry models.

\subsection{Object Efficiencies} 
\label{section:objeff}

The first method for deriving $A\epsilon$ to use in conjunction
with the model--independent limits is object efficiencies.
The person investigating a model would run a Monte Carlo generator
and place the acceptance cuts on the output which will determine 
the acceptance, $A$.  
The next step would
be to apply efficiencies (simple scale factors) 
for the identification of each object
in the signature, such as the photon or the $b$-quark tag.  
This has the advantage of being 
being very straightforward and the disadvantage that correlations
between the objects in the event are not accounted for.  For example,
a model with many jets would tend to have a lower efficiency for 
the photon isolation requirement than a model with few jets
and this effect would not be reflected in this estimate 
of the efficiency.

Using a sample of $Z\rightarrow e^+e^-$ events to measure the
efficiencies of the global event cuts, we find the $z<60$~cm cut is
92\% efficient.  The probability of finding no energy out-of-time is
98\%.  In this case the total global efficiency would be the product
of these two efficiencies.  In the discussion below, the efficiency of
the identification of each object is often listed as efficiencies of
several separate steps which should be multiplied to find the total
efficiency.

We can also use $Z\rightarrow e^+e^-$ events to measure the
efficiency of the photon identification cuts.  One electron from the
$Z$ is required to fire the trigger, but the second electron is
unbiased with respect to the trigger.  In addition the $Z$ peak
indicates the number of true physics events, ideal for measuring
efficiencies.  Which $Z$ electron is required to pass which set of
cuts (trigger or offline) must be effectively randomized to avoid
correlations between the two sets of cuts.  Requiring the cluster to
be far from the boundary of the active area in the calorimeters is
73\% efficient \cite{Toback thesis}.  The trigger is 91\% efficient,
the identification cuts are 86\% and the isolation requirement is 77\%
efficient.

For the $b$--quark efficiency we use a 70\% probability that the 
jets from the event are contained in the SVX. (This would be 64\% if the
global event vertex was not already required to have $z<60$~cm.)
We add a 90\% probability that the jet was taggable 
(containing two reconstructed tracks in the SVX, passing $p_t$ cuts) 
and apply the published \cite{Weiming Snowmass} tagging probability
as a function of the jet $E_t$ which can be summarized as
$$
\begin{array}{rcl}
0 &{\rm ~for~} &E_t<18~{\rm GeV} \\
0.35+0.00277*E_t &{\rm ~for~}& 18<E_t<72~{\rm GeV} \\
0.6  &{\rm ~for~}& E_t>72~{\rm GeV} \\
\end{array}
$$

The missing $E_t$ is found as the vector sum of the 
noninteracting  particles in the event.  As long as the missing $E_t$
is large, the resolution on the $\met$ should not greatly effect the 
efficiency.

In Section \ref{section:otherobs}, 
we searched the events in the $\gamma b$ sample for additional leptons; 
here we present approximate object efficiencies for those cuts.
These requirements and their efficiencies are borrowed from top--quark 
analyses\cite{top,tau}
as a representative selection for high--$E_t$ leptons.  The efficiencies quoted
here are measured in those contexts and therefore they are approximations
in a search for new physics.  In particular, the isolation efficiency
is likely to be dependent on the production model.  For example, 
if a model of new physics contained no jets, then the isolation efficiency
is likely to be greater than that measured in top--quark events
which contain several jets on average.

For the electron search we require $E_t>25$~GeV and $|\eta|<1.0$.
Given that an electron, as reported in the output of the 
Monte Carlo generator, passes these acceptance cuts, the probability 
that the 
electron strikes the calorimeter well away from any uninstrumented
region is 87\%, the probability to pass identification 
cuts is $80\%$, and to pass isolation cuts is approximately 
87\%\cite{Toback thesis}.  

For muons we require 
$p_t>25$~GeV and $|\eta|<$1.0.  
Given that the muon, as reported in the output of the
Monte Carlo generator,
passes these cuts, the fiducial acceptance of the muon detectors is
48\%.  Once the muon is accepted, the probability to pass identification 
cuts is $91\%$, and to pass isolation cuts is approximately 81\%.

Tau leptons are identified only in their one-- and three--prong 
hadronic decays which have a branching ratio of 65\%. (Tau 
semileptonic decays can contribute to the electron and muon searches.)
We require that the calorimeter cluster has $E_t>25$~GeV and $|\eta|<1.2$ and 
the object is not consistent with an electron or muon.
Given that the $\tau$ decays to a one-- or three--prong 
hadronic decay mode and passes the $E_t$ and $\eta$
requirements, the probability that the tau passes identification
and isolation cuts is approximately $57\%$.

In Section \ref{section:tests} we apply these object efficiencies to the 
supersymmetry models and compare to the results of the rigorously--derived
efficiency to test the accuracy of the results.

\subsection{Standard Model Calibration Process} 

The second method for determining $A\epsilon$ for a model
is the Standard Model process or efficiency model.  
In this method we select a simple physics
model that produces the signature.  The model is purely for the purpose
of transmitting information about 
$A\epsilon$ so it does not have any connection to 
a model of new physics.  Since it may be considered a calibration model,
it does not have to be tuned and will not become dated.  
The interested model builder runs a Monte Carlo of the new physics 
and places acceptance cuts on the output, determining $A$, the same
as the first step in the object efficiencies method.
This result is then multiplied by the 
value of $\epsilon$ which is taken to be the same as the 
value of $\epsilon$ which we report here for the standard model process.

We have adopted $WW$ production
as our efficiency model.  One $W$ is required
to decay to $e\nu$ and we replace the electron with a photon before the
detector simulation.  The second $W$ is forced to decay to $bu$,
so the combination yields the signature $\gamma b \met$.
Since some efficiencies may be dependent on the $E_t$ of the objects 
in the event, we will vary the ``$W$'' mass to present this effect.
A model--builder would then choose the efficiency that most closely
matches the mass scale of the new physics models.

\clearpage
The $A\epsilon$ for this model is found
using the same methods as used for the models of supersymmetry.
From the difference in the observed efficiencies in 
Monte Carlo and data $Z\rightarrow ee$ samples, we use a 14\% uncertainty 
on the efficiency of the photon ID and isolation.  
We use 9\% for the $b$--tagging efficiency uncertainty. 
The parton distribution function we use is CTEQ4L.  
Comparing this efficiency to those obtained
with the MRSD0$^\prime$ and GRV--94LO parton distribution functions,
we find a standard deviation of 5\%.
Turning off initial-- and final--state radiation increases the efficiency
by 2\% and 23\% respectively and we take half these numbers 
as the systematic uncertainty.
Varying the jet energy scale by 10\% causes the efficiency to change by 6\%.
We use an 4\% uncertainty for the luminosity.
In quadrature, the total systematic is 22\%.
Table \ref{tablimit} summarizes the results.

\begin{table}[ht]
\centering
\begin{tabular}{|c|c|c|c|c|c|} 
$M_W$ & 75~GeV & 100~GeV & 125~GeV & 150~GeV & 175~GeV \\ \hline
$\epsilon A$(\%)  & 0.85 & 2.56 & 4.86 & 6.98 & 8.12  \\ \hline
$\epsilon$  (\%)  & 11.8 & 10.7 & 13.9 & 15.4 & 16.5  \\ 
\end{tabular}
\caption{Summary of the efficiencies found for 
the values of $W$ mass used in the $WW$ calibration model
versus the value of the ``$W$'' mass.}
\label{tablimit}
\end{table}

In Section \ref{section:tests} we apply this method of 
calculating $A\epsilon$ to the 
supersymmetry models and compare to the results of the rigorously--derived
efficiency to test the accuracy of the results.

\subsection{A Public Monte Carlo} 

A Monte Carlo event generator 
followed by a detector simulation is the usual method 
for determining the efficiency of a model of new physics.
However, there is usually considerable detailed knowledge required
to run the simulation programs correctly so it is not practical to 
allow any interested person access to it.  But if the simulation 
is greatly simplified while still approximating the 
full program, it could become usable for any worker in the field.
The model--builder then only has to run this simple Monte Carlo
to determine $A\epsilon$.

An example of this kind of detector simulation, 
called SHW, was developed for the 
Fermilab Run II SUSY/Higgs Workshop \cite{workshop}.  
Generated particles are traced to 
a calorimeter and energy deposited according to a simple 
fractional acceptance and Gaussian resolution.  A list of tracks is
also created according to a simple efficiency and resolution model,
and similarly for muon identification.
The calorimeter energy is clustered to find jets.  Electromagnetic
clusters, together with requirements on isolation and tracking,
form electron and photon objects.  The tagging of $b$--quarks is
done with a simple, parameterized efficiency.
At points where object identification efficiencies would 
occur, such as a $\chi^2$ cut on an electron shower profile, 
the appropriate number of candidates are rejected to create 
the inefficiency.  The result is a simple list of objects that are
reconstructed for each event.
This method of determining efficiencies addresses the largest concern
not addressed in the previous methods --
the correlation of the characteristics of jets in the model
with isolation requirements.  We note that a highly parameterized
Monte Carlo has obvious limitations.

We have used the SHW program to compute efficiencies for the three 
models considered above.
Since the program is tuned to provide the approximate efficiency of the
Run II detector, we made a few minor changes to reflect the Run I 
detector.  In particular, we changed the photon fiducial efficiency
from 85\% to 73\% and the offline efficiency from 85\% to 60\%.
We reduced the SVX acceptance along the $z$ axis from 60~cm to 31~cm.
Finally, we removed soft lepton $b$-tagging and added a 90\% efficiency
for the global event cuts.

In the next section we use the public Monte Carlo to 
calculate $A\epsilon$ for the 
supersymmetry models and compare the results to the rigorously--derived
efficiency to test the accuracy of the public Monte Carlo.

\subsection{Tests of the Model--independent Efficiency Methods} 
\label{section:tests}

In the body of this paper we have provided rigorous limits on 
several variations of three supersymmetry physics models that produce 
the signature of $\gamma b \met$.  
In this section we apply the model-independent efficiency methods 
to the supersymmetry models.
We can then compare the results with the rigorous limits to 
evaluate how effective it is to apply the model--independent limits 
to real physics models.  

In most cases we need to distinguish between acceptance and
efficiency.  Acceptance, indicated by $A$, we define as the
probability for generated Monte Carlo objects to pass all geometric
and $E_t$ cuts.  For the $\gamma b\met$ signature, with $\met$ defined
as the vector sum of neutrinos and LSP's, the cuts defining the
acceptance of the signature are listed in Table
\ref{tab:acceptance}.

\begin{table}[ht]
\centering
\begin{tabular}{|c|c|c|} 
$\gamma$ & $E_t>25$~GeV & $|\eta|<$1.0 \\ \hline
$b$--quark & $E_t>30$~GeV & $|\eta|<$2.0 \\ \hline\hline
\multicolumn{3}{|c|}{Additional Signatures} \\ \hline
$\Sigma p_t(\nu,~LSP)$ & 
$\met >40$~GeV,$\Delta\phi(\gamma-\met)<168^\circ$ &  \\ \hline\hline
$e$ & $E_t>25$~GeV & $|\eta|<$1.0 \\ \hline
$\mu$ & $E_t>25$~GeV & $|\eta|<$1.0 \\ \hline
$\tau$ & $E_t>25$~GeV & $|\eta|<$1.2 \\ \hline
2$^{\rm nd}$ $\gamma$ & $E_t>25$~GeV & $|\eta|<$1.0 \\ \hline
2$^{\rm nd}$ $b$      & $E_t>30$~GeV & $|\eta|<$2.0 \\ \hline
Jets                  & $E_t>15$~GeV & $|\eta|<$2.0 \\ 
\end{tabular}
\caption{The list of requirements on the output of a Monte Carlo generator
which define the acceptance of a signature, $A$.  The requirements on the
photon and $b$--quark jet above the double line are common to all signatures
in this paper.  When missing $E_t$ is required, as in all the supersymmetry
searches and the tests of model--independent methods, both $\met>$40~GeV
and $\Delta\phi(\gamma-\met)<168^\circ$ are required.
The $\met$ requirement is removed and 
other requirements are added to make specific subsamples.}
\label{tab:acceptance}
\end{table}

\begin{figure}[!ht]
\epsfxsize=8.6cm
\hspace*{3.5cm}
\epsfbox{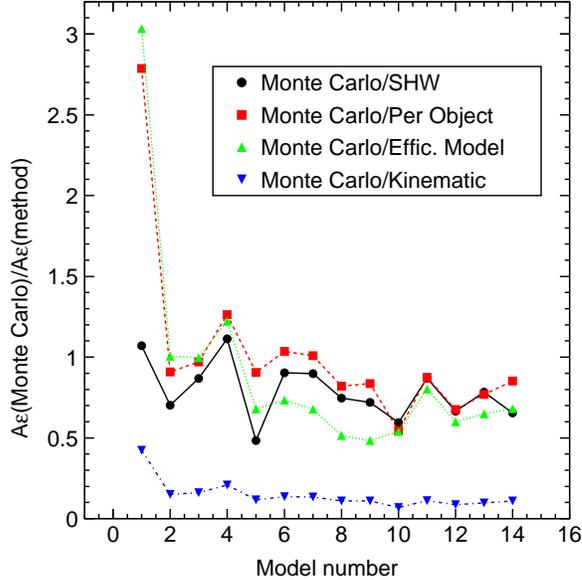}
\caption{The ratio of the efficiencies obtained with full detector 
simulation to those obtained with the model--independent methods.
The $x$ axis is the row number from 
Table \protect\ref{tab-object-results}.}
\label{fig-object-results}
\end{figure}

Table \ref{tab-object-results} and 
Figure \ref{fig-object-results} list the results.
The columns marked $R$ are the 
efficiency times acceptance done rigorously, divided by the 
the same found using each of the model--independent methods.
The difference of this ratio from 1.0 one is a measure of
the accuracy of the approximate methods compared to the rigorous method.

\begin{table}[ht]
\centering
\begin{tabular}{|c|c|c|c|c|c|c|c|} 
Model   &$M_s$& BR(\%) & $A$  & $A\cdot\epsilon$ 
& $R_{obj}$  & $R_{WW}$  & $R_{SHW}$  \\ \hline
   &130&   3&   65.0&  27.50&   2.79&   3.03&   1.07 \\ \cline{2-8}
GMSB
   &147&  20&   49.8&   7.45&   0.91&   1.00&   0.70 \\ \cline{2-8}
$M_s = M_{{{\tilde\chi}_1^\pm}}$
   &170&  23&   51.7&   8.35&   0.97&   1.00&   0.87 \\ \cline{2-8}
   &186&  18&   54.7&  11.44&   1.26&   1.22&   1.11 \\ \hline\hline
   &185&  30&   17.0&   1.97&   0.91&   0.68&   0.48 \\ \cline{2-8}
${{\tilde\chi}^0_2}\rightarrow\gamma {{\tilde\chi}^0_1}$
   &210&  30&   22.0&   2.98&   1.04&   0.73&   0.90 \\ \cline{2-8}
$\tilde q$, $\tilde g$ production
   &235&  30&   24.0&   3.23&   1.01&   0.68&   0.90 \\ \cline{2-8}
$M_s = M_{\tilde g}$
   &260&  30&   24.5&   2.69&   0.82&   0.52&   0.75 \\ \cline{2-8}
   &285&  30&   19.7&   2.16&   0.84&   0.48&   0.72 \\ \hline\hline
   &110& 100&   13.5&   0.93&   0.54&   0.54&   0.59 \\ \cline{2-8}
${{\tilde\chi}^0_2}\rightarrow\gamma {{\tilde\chi}^0_1}$
   &130& 100&   12.6&   1.41&   0.88&   0.80&   0.87 \\ \cline{2-8}
$\tilde q$, $\tilde g$ production
   &140& 100&   14.8&   1.29&   0.68&   0.60&   0.66 \\ \cline{2-8}
$M_s = M_{{{\tilde\chi}_2^\pm}}$
   &150& 100&   13.7&   1.34&   0.77&   0.65&   0.78 \\ \cline{2-8}
   &170& 100&   11.5&   1.27&   0.85&   0.68&   0.65 \\ 
\end{tabular}
\caption{The results of comparing the methods of calculating $A\epsilon$
using the model--independent methods and the rigorously--derived $A\epsilon$.
Each row is a variation of a model of supersymmetry as indicated
by the label in the first column and the mass of a supersymmetric
particle listed in column two (GeV).  The column labeled $A$ is the 
acceptance of the model in \% and the next column is the rigorously--derived 
$A\epsilon$.  The columns labeled with $R$ are the ratios of 
the rigorously--derived $A\epsilon$ to $A\epsilon$ found using the 
model--independent method indicated.}
\label{tab-object-results}
\end{table}

\subsection{Conclusions from tests} 

There are several notable effects apparent immediately from 
Table \ref{tab-object-results} and 
Figure \ref{fig-object-results}.
The first is that the comparison of efficiencies for one model
point fares especially poorly.  This occurs when the branching
ratio for the model is very small (2\%).  When the events do not
contain many real photons and $b$ quarks, the small number of objects
misidentified as photons and $b$-quarks becomes important.  For example, 
jets may pass photon cuts and $c$ quarks may be $b$-tagged.
When this occurs, the full simulation will be more efficient than
a method which specifically requires that the Monte Carlo generate an 
isolated 
photon or $b$-quark in order to accept the event.  This is true of the 
object efficiencies method and the efficiency model method.  We note 
that the public Monte Carlo method does allow misidentification 
and so it does not
show this large mismatch.  We can conclude that when the branching ratios are
small, the public Monte Carlo method is vastly superior to the 
others.

In the object efficiency method, the acceptance of the signature is 
computed by running the Monte Carlo without a detector simulation.  
As each object in the signature is identified and passes acceptance cuts, 
the individual object
efficiencies are multiplied.  
These object efficiencies which may or may not 
be $E_t$-- or $\eta$--dependent, are listed in Section 
\ref{section:objeff}.  In this test, these 
efficiencies are typically well--matched to the rigorously--derived 
efficiencies.
The average of $R_{obj}$ over all models except the first, is $0.88\pm 0.21$
where 0.21 is the RMS computed with respect to 1.0, 
the ideal result.

In the efficiency model method, we generate a Monte Carlo model that
is not related to a search for new physics but produces the signature
of interest.
For the signature of $\gamma b \met$, we generated 
$WW\rightarrow (\gamma\nu)(bu)$. 
The efficiency model results are also optimistic, the average is 
a ratio of $0.74\pm 0.35$ where again the uncertainty is actually the RMS  
with 1.0, the ideal result.  We found that the difficulty of applying
this method was in choosing the mass scale.  For example, we chose to 
match the ``$W$'' mass to the ${{\tilde\chi}_2^\pm}$ mass in the direct 
production of
the ${{\tilde\chi}^0_2}\rightarrow\gamma {{\tilde\chi}^0_1}$ model.  
However, the photon
comes from a secondary decay and the effect of 
the LSP mass compared to the massless neutrino causes the $E_t$ of
the $\gamma$ and $b$ to be poorly matched to the $E_t$ of
these objects in the $WW$ model.

In the public Monte Carlo method, we compute the efficiency using 
SHW, a highly-parameterized, self--contained Monte Carlo.
In general, results here are somewhat optimistic
with the average ratio to the rigorous total efficiency being 
$0.77\pm 0.28$, where the uncertainty is the RMS 
computed with respect to 1.0, the ideal result.
                                             
For completeness we also include the ratio of the simple acceptance
to the rigorous acceptance times efficiency.  The average ratio
is $0.12\pm 0.87$.  

The methods for calculating efficiency without access to the full
detector simulation are accurate to approximately 30\% overall.  They
tend to underestimate $A\epsilon$ by 10-25\% but the result for
individual comparisons varies greatly.  These uncertainties are larger
than, but not greatly larger than, a typical uncertainty in a
rigorously--derived efficiency, which is 20\%.

We conclude that to determine if a model is easily excluded or 
far from being excluded by the data, the model--independent
methods are sufficient.  If the model is within 30\% of exclusion,
no conclusion can be drawn and the efficiency should be rigorously--derived.



\begin{thebibliography}{99}
\bibitem{coo} The CDF coordinate system is right-handed, and has the $z$ axis
aligned with the proton beam and $y$ vertical. The azimuthal angle is denoted
as $\phi$ and the
polar angle as $\theta$; $\eta$ is $-\ln[\tan(\theta/2)]$.
\bibitem{uncert}
In this Letter when two uncertainties are reported, the first is
statistical and the second is systematic.
\end{thebibliography}
\end{document}